%% file: main.tex
\documentclass[sigconf, screen,nonacm]{acmart}
\acmConference[EASE 2025]{The 29th International Conference on Evaluation and Assessment in Software Engineering}{17–20 June, 2025}{Istanbul, Türkiye}
\usepackage{mathtools}
\usepackage{algorithmic}
\usepackage{graphicx}
\usepackage{textcomp}
\usepackage{xcolor}
\usepackage{dsfont}
\usepackage{cleveref}
\usepackage{booktabs}
\usepackage{subcaption}
\usepackage{tcolorbox}
\usepackage{multirow}
\usepackage{paralist}
\usepackage{enumitem}
\usepackage{makecell}
\usepackage{multirow}
\usepackage{nicefrac}
\usepackage{xspace}

\definecolor{RED}{HTML}{DC3220}
\definecolor{BLUE}{HTML}{005AB5}
\definecolor{GREEN}{HTML}{009E73}

\newcommand{\var}[1]{\texttt{\mbox{#1}}}

\usepackage{tikz}
\usetikzlibrary{arrows, positioning, shapes, calc, fit}
\tikzset{
    vertex/.style={ellipse,draw, thick, inner sep=0, minimum height=1.5em},
    edge/.style={->,> = latex', thick}
}

\theoremstyle{definition}
\newtheorem{definition}{Definition}

\begin{document}

\title{Using Causal Inference to Test Systems with Hidden and Interacting Variables: An Evaluative Case Study}
\thanks{Foster, Walkinshaw, Hierons, and Wild were supported by EPSRC CITCoM grant [EP/T030526/1]. Donghwan Shin was supported by EPSRC SimpliFaiS grant [EP/Y014219/1].
This is a preprint. The full version of the paper will appear in the proceedings of EASE 2025.}

\author{Michael Foster}
\email{m.foster@sheffield.ac.uk}
\orcid{0000-0001-8233-9873}
\affiliation{%
  \institution{University of Sheffield}
  \city{Sheffield}
  \country{UK}
}\author{Robert M. Hierons}
\email{r.hierons@sheffield.ac.uk}
\orcid{0000-0002-4771-1446}
\affiliation{%
  \institution{University of Sheffield}
  \city{Sheffield}
  \country{UK}
}
\author{Donghwan Shin}
\email{d.shin@sheffield.ac.uk}
\orcid{0000-0002-0840-6449}
\affiliation{%
  \institution{University of Sheffield}
  \city{Sheffield}
  \country{UK}
}
\author{Neil Walkinshaw}
\email{n.walkinshaw@sheffield.ac.uk}
\orcid{0000-0003-2134-6548}
\affiliation{%
  \institution{University of Sheffield}
  \city{Sheffield}
  \country{UK}
}
\author{Christopher Wild}
\email{c.wild@sheffield.ac.uk}
\orcid{0009-0009-1195-1497}
\affiliation{%
  \institution{University of Sheffield}
  \city{Sheffield}
  \country{UK}
}

\begin{abstract}
Software systems with large parameter spaces, nondeterminism and high computational cost are challenging to test.
Recently, software testing techniques based on causal inference have been successfully applied to systems that exhibit such characteristics, including scientific models and autonomous driving systems.
One significant limitation is that these are restricted to test properties where all of the variables involved can be observed and where there are no interactions between variables. In practice, this is rarely guaranteed; the logging infrastructure may not be available to record all of the necessary runtime variable values, and it can often be the case that an output of the system can be affected by complex interactions between variables.
To address this, we leverage two additional concepts from causal inference, namely effect modification and instrumental variable methods.
We build these concepts into an existing causal testing tool and conduct an evaluative case study which uses the concepts to test three system-level requirements of CARLA, a high-fidelity driving simulator widely used in autonomous vehicle development and testing.
The results show that we can obtain reliable test outcomes without requiring large amounts of highly controlled test data or instrumentation of the code, even when variables interact with each other and are not recorded in the test data.
\end{abstract}




\maketitle

\input{sections/01-introduction}
\input{sections/02a-motivation}
\input{sections/02b-causal-testing}
\input{sections/03-research-design}

\input{sections/04-studied-cases}
\input{sections/05-results}
\input{sections/07-validity-threats}
\input{sections/08-related-work}
\input{sections/09-conclusion}

\bibliographystyle{ACM-Reference-Format}
\bibliography{main}

\end{document}

%% file: sections/01-introduction.tex
\section{Introduction}\label{sec:introduction}
Complex software systems appear in a broad range of applications such as computational models, autonomous driving systems, and cyber-physical systems.
These have several fundamental characteristics that make them difficult to test, notably large input spaces, nondeterminism, and uncontrollable behaviour.
Furthermore, long runtimes and high computational cost often limit the number of tests that can be feasibly executed.

Causal reasoning is increasingly being applied to address these testing challenges \cite{clark2022ctf,poskitt2023,giamattei2024causality}.
Well-established in fields such as epidemiology and sociology, the idea is to specify a model of the expected causal relationships between variables and use this to identify and remove bias when applying statistical estimation techniques.
This enables the expected causal effects to be validated using pre-existing uncontrolled data rather than requiring a specially curated dataset, without risking test outcomes being made unreliable by a biased data generation process.
% These techniques make testing more efficient by enabling the use of pre-existing runtime data, rather than requiring specially curated test data, while still yielding reliable test outcomes.

However, there are two essential challenges that have not been considered by previous work on causality-based testing:
\begin{inparaenum}[(1)]
  \item \label{ch:observable} Non-observability: Key variables that are required to evaluate correctness properties may not be observable.
    When testing relationships between software inputs and outputs, we need to account for other inputs and internal variables to isolate the causal effect of interest.
    If any of those variables are missing from the test data (e.g. if they are not logged during execution), this can lead to biased and unreliable test outcomes.
  \item \label{ch:interaction} Interacting variables: Software outputs may depend on \emph{combinations} of interacting variable values.
  So far, causal relationships between variables have only been considered in isolation, neglecting faults involving interactions.
\end{inparaenum}

We perform an evaluative case study \cite{ralph2021empirical} to investigate how two additional concepts from causal inference --- effect modification and instrumental variable methods \cite{hernan2020causal} --- can address these challenges.
Effect modification allows us to reason about the causal effects of interacting variables.
Instrumental variable methods allow us to adjust for variables that are missing from the test data.
While these are both well-established causal inference techniques, this is the first work to explicitly apply them to a software engineering context.
Our main contributions are as follows:
\begin{itemize}[leftmargin=1em,nosep]
    \item We apply effect modification from causal inference to reason about interactions between variables when testing software.
    \item We apply instrumental variable methods from causal inference to reason about unobservable variables when testing software.
    \item We perform an evaluative case study considering three testing requirements in the context of the CARLA high-fidelity driving simulator~\cite{dosovitskiy2017carla}.
    The results show that the above techniques can yield reliable test outcomes for software with interacting and unobservable variables.
\end{itemize}

The remainder of this paper is structured as follows.
\Cref{sec:movivating-example} introduces the CARLA simulator and the testing challenges we consider in this work.
\Cref{sec:causal-testing} gives background on causal software testing and the essential elements of causal inference that we use in this work. For a more comprehensive introduction, we refer the reader to \cite{pearl2009causality,hernan2020causal}.
\Cref{sec:research-design} lays out the design of our evaluative case study.
\Cref{sec:studied-cases} shows how we used causal inference to test our three requirements.
\Cref{sec:results} provides answers to our research questions.
\Cref{sec:validity-threats} discusses potential threats to validity and our chosen mitigation strategies.
\Cref{sec:related-work} highlights key related works.
Finally, \Cref{sec:conclusion} concludes the paper.

%% file: sections/02a-motivation.tex
\section{Testing Challenges}\label{sec:movivating-example}
In this section, we outline the main testing challenges considered in this paper in the context of a motivating example concerning automated driving system (ADS) testing. While the testing challenges we consider in this work are not unique to the field of ADS testing, they are particularly pronounced here, which makes it an ideal context within which to explore our research questions.

\subsection{Motivating Example: CARLA Driving Simulator}
CARLA \cite{dosovitskiy2017carla} is a popular open-source high-fidelity driving simulator developed to support the development, training, and validation of ADSs.
The CARLA GitHub repository \cite{carlagithub} has over 10,000 stars and over 3,000 forks at the time of writing.
CARLA provides a wide range of configurable driving scenario entities, such as weather conditions, traffic lights, non-playing character (NPC) vehicles (i.e. traffic), and pedestrians. This makes CARLA the state-of-the-art ``system'' \cite{kaur2021simulators} for simulation-based ADS testing.

The CARLA leaderboard \cite{carlaleaderboard} is a benchmark for evaluating ADSs.
Its V1.0 SENSOR track, which restricts ADSs to sensor inputs (e.g. cameras), has 36 entrants at the time of writing.
ADSs are scored on their ability to drive predefined driving scenarios from start to end in a given time.
Penalties are applied for infractions, such as collisions or running red lights.
To enable a fair comparison, these penalties must be implemented correctly.
If CARLA were a conventional software system, this would be trivial to test: we would simply commit each infraction and check that the correct penalty was applied.
Unfortunately, CARLA exhibits four characteristics that make this impractical -- nondeterminism, limited observability, interactions between parameters, and long runtimes.

\begin{description}[nosep, wide=0pt]
\item[Nondeterminism]
CARLA can produce different behaviours for different runs of the same input configuration.
For example, pedestrian movement is completely random, even for the same seed \cite{pedestrianissue}.
Since CARLA cannot spawn an agent if their spawn point is occupied, we may end up with fewer pedestrians than specified if they move into each other's spawn points.
This means that the impact of particular configurations can only be studied statistically using multiple runs.
Nondeterminism also raises issues of controllability \cite{freedman1991testability}: we cannot reliably elicit particular behaviours.
This makes it hard to isolate the effect of any particular input.

\item[Observability]
CARLA has a large number of configuration parameters and internal variables, many of which are not logged by default.
This means that, not only are we unable to \emph{control} certain aspects of the simulation directly, but we cannot even \emph{observe} them.
For example, CARLA does not record how many pedestrians and NPC vehicles were successfully spawned into the simulation.
Again, this makes it hard to isolate the effects of particular inputs.
% For example, if we were to introduce a new feature to CARLA and wanted to test that it did not affect the runtime, we would need to repeatedly run the two versions under the same conditions.
% We clearly cannot do this if we do not know what the actual simulation conditions were.

\item[Interaction]
Much of CARLA's behaviour depends on complex interactions between parameters.
% For example, the weather and the numbers of pedestrians and NPC vehicles will all affect how long it takes to drive a particular scenario, and the effect of changing any one of these may depend on the values of the others.
For example, the numbers of pedestrians and NPC vehicles both affect how long it takes to drive a particular scenario, with the delay caused by increasing the number of pedestrians being compounded by busy roads, since more vehicles will have to stop to allow pedestrians to cross, causing longer traffic jams and greater disruption.
This interaction further adds to the difficulty in isolating the effect of any individual input.

\item[Execution Time and Computational Cost]
CARLA requires high-end hardware \cite{carlagithub}, and is time-consuming and computationally demanding to execute.
Furthermore, CARLA tends to run slower than real time: it takes longer than 1 second of real-world time to run 1 second of the simulation.
For example, the driving scenarios we collected as part of this study took an average of around 13 minutes to execute.
The longest took over two hours, even though it was still only a few minutes of simulation time.
% For example, \citet{nitsche2018junctions} took around 142 hours to test a single driving scenario consisting of a right turn at a simple T-junction.
Thus, a tester can only consider a small fraction of potential test executions, especially if configurations need to be run repeatedly to mitigate nondeterminism.
Since the events that are of interest from a testing perspective (e.g., collisions) tend to occur relatively rarely, a premium is placed on the ability to extrapolate as much useful information as possible from the few test executions that can be collected.
\end{description}

\subsection{Limitations of Existing Techniques}\label{sec:limitations}
Within the research literature on ADS testing, the notion of faulty behaviour is often restricted to faults that are easily detectable, such as obvious driving violations (e.g. collisions or running a red light) \cite{zhong2021survey,sun2022scenario,zhang2023finding}. The authors are not aware of any testing approaches that can test behaviour against more nuanced requirements (e.g. ``The model of ego-vehicle should not affect how often it crashes.''). We suspect that this is at least partially because of the practical challenges that this would entail (as mentioned above).

In principle, statistical metamorphic testing (SMT) \cite{guderlei2007statistical} provides a framework within which to test such properties. The SMT approach involves repeatedly running the software under two (or more) configurations and performing statistical tests on the resulting output data.
For example, we would run the ADS several times with two different ego-vehicles, and perform a hypothesis test to investigate whether either ego-vehicle had significantly more crashes.
This is similar to A/B testing \cite{siroker2015abtesting}, where different groups of users are assigned different versions of software to see which performs better.
The main limitation of these approaches is that all variables must be carefully controlled in the manner of a laboratory experiment.
This may not always be possible, especially when testing relationships between different software outputs.
Furthermore, test data must be collected separately for each property being tested, meaning that large amounts of test data are often required.

%% file: sections/02b-causal-testing.tex
\section{Causal Testing}\label{sec:causal-testing}
Several recent techniques apply the model-based statistical framework of causal inference (CI) \cite{pearl2009causality} to test software.
This \emph{Causal Testing} excels for testing properties of nondeterministic systems where it is difficult to obtain large numbers of carefully controlled executions, such as computational models \cite{clark2022ctf} and ADSs \cite{giamattei2024causality}.

As with SMT, multiple runs of the software are used to draw statistical conclusions about the \emph{relationships} between program inputs, outputs, and internal variables.
However, CI explicitly separates the \emph{collection} and \emph{analysis} of test data by employing domain knowledge supplied by the tester in the form of a causal model that specifies the expected causal relationships between program variables.
This means that the test cases can be evaluated using pre-existing runtime data rather than specially curated test data.

% \citet{clark2022ctf} introduce the term Causal Testing, which we will also use throughout this work.
\newcommand{\CTstepDAG}{Specify the Causal Model\xspace{}}
\newcommand{\CTstepCases}{Define Causal Test Cases\xspace{}}
\newcommand{\CTstepData}{Collect Test Data\xspace{}}
\newcommand{\CTstepEvaluate}{Evaluate the Causal Test Cases\xspace{}}
Causal Testing applies to properties framed as the effect of a treatment on an outcome, and has four main steps:
\begin{inparaenum}[(1)]
  \item\label{CTstepDAG} \CTstepDAG,
  \item\label{CTstepData} \CTstepData,
  \item\label{CTstepCases} \CTstepCases, and
  \item\label{CTstepEvaluate} \CTstepEvaluate.
\end{inparaenum}
These are elaborated in the following sub-sections.
In principle, everything except the initial formation of the causal model can be (semi-)automated.

\subsection{\CTstepDAG}
The first step is to specify the expected causal relationships between variables in the system using a directed acyclic graph (DAG), exemplified in \Cref{fig:iv}.
Nodes represent variables, and an edge $X \to Y$ represents the domain knowledge that $X$ may have a direct causal effect on $Y$.
The absence of such an edge means that $X$ \emph{definitely does not} have a direct causal effect on $Y$.
A causal DAG should include all inputs, outputs, and internal variables that are relevant to the properties being tested, even if they cannot be controlled or observed.
By analysing paths in a DAG \cite{pearl2009causality}, it is possible to identify which variables must be adjusted (controlled for) to isolate the causal effect of $X$ on $Y$.
We provide an example in \Cref{sec:studied-cases}.

Causal DAGs form an intuitive model of the system under test and are widely used in fields such as epidemiology and sociology \cite{hernan2020causal}, where they are often hand-drawn by domain experts.
As with any model-based testing technique, drawing a DAG requires domain knowledge since it forms part of the test oracle \cite{barr2015oracle}.
However, causal DAGs are much lighter weight than traditional models, such as finite state machines \cite{chen1995efsm}.
They do not specify the precise form of the relationships between variables, merely their existence.

\subsection{\CTstepData}
The second step is to collect test data.
A major benefit of Causal Testing is that this data can be ``observational'', i.e., collected without needing to tightly control the inputs.
The advantage of this from a software engineering standpoint is that the same test data can be reused to test multiple properties \cite{clark2022ctf},
without requiring carefully controlled test data generation.
% without imposing strong assumptions as to how the data was generated.
For example, it would be valid to use pre-existing log data recorded during normal use.
The important limitation is that the test data must satisfy the \emph{positivity} assumption, which is fundamental to CI \cite{hernan2020causal}.
Formally, this means that the probability of each treatment (typically an input configuration) of interest must be non-zero.
Intuitively, this means that test outcomes are more accurate and reliable if the test data achieves a good coverage of the input space.

\subsection{\CTstepCases}
The next step is to encode the properties to be tested as causal test cases.
These are intuitively similar to metamorphic relations \cite{chen1998metamorphic} in that we observe the effect of \emph{changing} a particular variable.
\Cref{def:causal-test-case} formalizes this, and is slightly adapted from \cite{clark2022ctf}.

\begin{definition}\label{def:causal-test-case}
    Given a causal DAG $G$ representing the expected causal relationships between variables of the system under test, a \emph{causal test case} is a triple $(X, Y, E)$, where $X$ and $Y$ are nodes in $G$, respectively referred to as the \emph{treatment} and \emph{outcome}.
    $E$ is the expected causal effect of $X$ on $Y$, serving as the test oracle \cite{barr2015oracle}.
\end{definition}

For example, to test that the model of the ego-vehicle does not affect the number of infractions it commits, we would define our treatment variable $X$ to be the model of the car, the outcome $Y$ to be the number of infractions, and the expected causal effect $E$ of $X$ on $Y$ to be zero (indicating no effect).

% In principle, a causal DAG can be automatically transformed into a suite of causal tests \cite{clark2022ctf} by specifying a causal test case for each pair of nodes $(X, Y)$ in the DAG.
% The expected \emph{direct} causal effect is then non-zero (i.e. \emph{some effect}) if there exists an edge $X \to Y$ in the DAG and zero (i.e. \emph{no effect}) if there does not.
% Where there exists a path, but no edge, from $X \to \ldots \to Y$, this is tested by the individual edges on the path.
% In this way, the causal DAG can serve as a test oracle in itself, but the expected causal relationships can be made more precise if more about their nature is known, e.g. a positive or negative relationship, or even a specific value or range of values.

\subsection{\CTstepEvaluate}\label{sec:causal-testing:eval}
The final step is to use CI to evaluate each test case.
There are three sub-steps to this.

\textbf{Identification:}
First, the DAG is used to identify which variables need to be adjusted to remove bias.
This is done by automatically searching for ``backdoor paths'' in the DAG and variables which can be controlled to close them \cite{pearl2009causality}.
A common source of bias is \emph{confounding}, where a third variable $Z$ has causal paths to both the treatment $X$ and the outcome $Y$, which can introduce a spurious correlation between $X$ and $Y$, even if there is no direct causal link between them.
To adjust for this, $Z$ (and other sources of bias) are controlled for by including them as features in the estimation step (below) so that their values are properly taken into account.

\textbf{Estimation:}
Next, we use the test data to estimate the causal effect, with 95\% confidence intervals \cite{o2016interpret}.
This involves using a statistical estimator, such as regression.
% Rather than using all of the features in the data, as traditional machine learning approaches might do, causal approaches use only the treatment variable $X$ and the set of variables identified by the identification step.
In this work, we estimate unit Average Treatment Effect (ATE), which represents the change in the outcome $Y$ we would expect to see if we increased the treatment $X$ by 1.
In a linear setting with $Y = a X + b Z + c$ (where $a$, $b$, and $c$ are constant coefficients), this is given by $a$.
To test non-linear relationships, one can add extra terms (e.g. powers, reciprocals, interaction terms) or use a machine learning model \cite{mcConnell2019estimating} if the equational form is not known.
The advantage of CI is that the identification step automatically identifies the relevant features from the DAG instead of factoring in all of the (potentially irrelevant) features in the data.

\textbf{Comparison:} Finally, the causal effect estimate is checked against the expected causal effect $E$ to determine the test outcome.
At the coarsest level, we can simply check for the presence or absence of a causal effect.
For unit ATE (defined above), there is deemed to be a causal effect if the confidence intervals do not contain zero.
A causal DAG can act as a test oracle in itself, and can be automatically transformed into a suite of causal tests that validate the specified causal effects and independence relations \cite{clark2023metamorphic}.
It is also possible to make the test oracle more precise by, for example, checking for a positive or negative causal effect, or even a specific value.

\subsection{Handling Interaction and Unobservable Variables}
Previous Causal Testing research has been limited to the analysis of causal effects between pairs of variables in systems where all relevant variable values are recorded in the test data.
However, two additional challenges still present a barrier to its broader application: interaction and unobservable variables.
% Effect modification is alternative source of bias that can still effect causal test estimates and the reliability of test outcomes even after confounding has been adjusted for.
% Unobservable variables do not appear in software logs, but still need adjusting for if they confound a causal effect.
% Instrumental variable methods from CI provide a potential solution to this, without having to implement additional logging.
% In \Cref{sec:studied-cases}, we investigate instrumental variable methods from CI as a potential solution for this.

\subsubsection{Interaction and Effect Modification}\label{sec:effect-modification}
Many software faults manifest themselves as interactions between multiple variables \cite{nie2011combinatorial}, where several variables may need to take particular values.
When this is the case, the causal effect of one variable on the outcome is modified by another variable.
In CI, this is known as \emph{effect modification} \cite{hernan2020causal}: the actual \emph{relationship} $X \to Y$ is changed, depending on the value of a third variable $Z$.
The CI solution to this is to include an \emph{interaction term} as an additional feature when estimating causal effects.
For example, we may use the regression equation $Y = aX + bZ + cXZ + d$, where $XZ$ is the interaction term.
In \Cref{sec:studied-cases}, we investigate whether this allows us to obtain reliable test outcomes when variables interact.
% For example, passing an audio signal through a device that doubles the volume would effectively double the effect of each volume increment.
% This is different from simply ``stacking '' the causal effects of different features.
% For example, passing an audio signal through a device which reduces the volume by 25dB would not change the volume \emph{increments}, merely the final output volume.
% In \Cref{sec:infraction-penalties}, we investigate how Causal Testing can tackle this.
% \MF{It would be cool if we could think of a more software engineering-y example, but I'm struggling to think of anything that's this intuitive.}

% Unlike direct causal effects, there is no standard notation for depicting effect modification on a causal DAG.
% In this paper, we use the notation proposed in \cite{weinberg2007modification} of drawing edges from nodes to other edges, as can be seen in in \Cref{fig:dag}.
% As with confounding variables, effect modifiers can be adjusted for by controlling for the variables when estimating the causal effect.

\subsubsection{Unobservable Variables}\label{sec:instrumental-variables}
Software logs may be incomplete \cite{chen2021logs} and may not record every variable during execution.
This can lead to biased, unreliable test outcomes as we may not be able to adjust for confounding variables by controlling their values.
While we can sometimes instrument programs to provide extra logging, this may not always be possible.

\begin{figure}[ht!]
    \centering
    \input{figures/dag-iv}
    \caption{General setup for IVs. The unobserved variable \var{U} is highlighted in gray.}
    \Description{General setup for IVs. The unobserved variable \var{U} is highlighted in gray.}
    \label{fig:iv}
\end{figure}
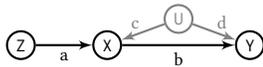

Instrumental variable (IV) methods \cite{pearl2009causality,wright1920relative} from CI provide an elegant solution to this problem under certain circumstances.
As an example, consider the causal DAG in \Cref{fig:iv}.
This shows the causal relationships between four variables: \var{X}, \var{Y}, \var{Z}, and \var{U} (which is unobserved), along with path coefficients \cite{wright1920relative} ($a$, $b$, $c$, $d$) that represent the unit ATE of each causal relationship.
To estimate the direct effect of \var{X} on \var{Y} ($b$ in \Cref{fig:iv}) in the presence of confounder \var{U}, we would typically need data for \var{U} to adjust its biasing effect.

Instead, we can use \var{Z} as an \emph{instrument} to calculate $b$ without needing data for \var{U}.
To do this, we divide the total effect of \var{Z} on \var{Y} ($ab$) by the direct effect of \var{Z} on \var{X} ($a$).
That is, $\nicefrac{ab}{a} = b$.
This is possible because \Cref{fig:iv} satisfies the following three conditions and operates in a linear setting:
\begin{inparaenum}[(1)]
    \item there is no arrow between \var{U} and \var{Z},
    \item there is an arrow between \var{Z} and \var{X}, and
    \item there is no direct arrow from \var{Z} to \var{Y}.
\end{inparaenum}
Although these assumptions can be restrictive, and IV methods tend to give less precise estimates \cite{hernan2020causal}, they nevertheless enable us to draw causal conclusions about relationships between variables, which would otherwise be impossible.
In \Cref{sec:carla-version}, we investigate whether IV methods allow us to obtain reliable test outcomes for systems with unobservable variables but where the DAG conforms to the above constraints.

%% file: figures/dag-iv.tex
 \begin{tikzpicture}[scale=1]
    \footnotesize
    \node[vertex, circle] (Z) at (0, 0) {\var{Z}};
    \node[vertex, circle, align=center, anchor=west, shift={(0.75, 0)}] (X) at (Z.east) {\var{X}};
    \node[vertex, circle, align=center, anchor=west, shift={(1.5, 0)}] (Y) at (X.east) {\var{Y}};
    \node[gray, vertex, circle, anchor=south, shift={(0, 0.15)}] (U) at ($ (X) !.5! (Y) $) {\var{U}};

    \draw[edge] (Z) -- (X) node[midway, below] {a};
    \draw[edge] (X) -- (Y) node[midway, below] {b};
    \draw[gray, edge] (U) -- (X) node[midway, above, shift={(-1mm, -0.5mm)}] {c};
    \draw[gray, edge] (U) -- (Y) node[midway, above, shift={(1mm, -0.5mm)}] {d};
\end{tikzpicture}

%% file: sections/03-research-design.tex
\section{Research Design}\label{sec:research-design}
We perform an evaluative case study \cite{ralph2021empirical} to gain an in-depth understanding of \emph{how} Causal Testing applies to software systems like CARLA, where interacting and unobservable variables prevent us from obtaining reliable test outcomes with current techniques.
The case study methodology is well suited to this because our object of study is a contemporary phenomenon (Causal Testing) that must be studied in its context (by testing a system) and not in isolation \cite{runeson2012case}.
Our study design follows the guidelines of \citet{runeson2012case}.

\subsection{Objectives and Research Questions}
The goal of this study, stated using the Goal-Question-Metric (GQM) approach \cite{basili1994encyclopedia}, is to
``analyse Causal Testing \emph{for the purpose of} evaluation and characterisation of its testing ability \emph{with respect to} software systems with nondeterminism and limited controllability and observability \emph{from the point of view of} software testers \emph{in the context of} ADS testing''.
Critically, we are investigating the \emph{process} of Causal Testing rather than trying to find faults in individual systems.
Furthermore, the application of Causal Testing to CARLA is not about generating test scenarios, like many existing ADS testing studies~\cite{zhong2021survey,sun2022scenario,zhang2023finding}.
Causal Testing is not intended to replace the existing ADS testing techniques but should instead be considered complementary.
We achieve our goal by answering the following research questions.
\begin{enumerate}[label=\textbf{RQ\arabic*},ref=\arabic*,wide, labelwidth=0pt, labelindent=0pt]
    \item\label{rq:controlability:parameters} \textit{Can Causal Testing deliver reliable test outcomes for software with interacting parameters?}
    To answer RQ\ref{rq:controlability:parameters} we examine whether the use of interaction terms produces more reliable test outcomes.

    \item\label{rq:controlability:data} \textit{Can Causal Testing deliver reliable test outcomes when using uncontrolled data?}
    To answer RQ\ref{rq:controlability:data}, we compare causal effect estimates calculated using SMT-style data with those calculated using a smaller amount of less controlled data.

    \item\label{rq:observability} \textit{Can Causal Testing deliver reliable test outcomes for software with unobservable parameters?}
    To answer RQ\ref{rq:observability}, we compare causal effect estimates calculated using IV methods, traditional adjustment (which requires the values to be observed), and no adjustment to investigate how this affects the \emph{accuracy} of our estimates and the \emph{reliability} of test outcomes.

    \item\label{rq:faults} \textit{Can Causal Testing discover faults under the above circumstances?}
    To answer RQ\ref{rq:faults}, we consider the unexpected behaviour we encountered when testing our requirements.
\end{enumerate}

\subsection{Case Selection and Units of Analysis}
Our case study is charactesized as \emph{single-case} and \emph{embedded} \cite{runeson2012case}.
Our case is the CARLA platform, and we have multiple units of analysis embedded within this.
Our units of analysis are the following three requirements, which we selected to enable us to investigate key aspects of Causal Testing and answer our RQs, while still being relevant to ADS testing.

% [label=\textbf{RE\arabic*},ref=RE\arabic*,wide, labelwidth=!, labelindent=0pt]
% [\bf RE1]
\begin{enumerate}[label=\textbf{RE\arabic*},ref=RE\arabic*,wide,labelwidth=0pt,labelindent=0pt]
    \item \label{req:infraction-penalties} (Infraction penalties).
        The CARLA leaderboard evaluates an ADS's ability to drive a set of predefined scenarios from start to end in a given time.
        The \var{DrivingScore} is then calculated as the proportion of the route that the ADS managed to complete within its lane, with penalties being applied for any infractions committed.
        This is shown in \Cref{eq:dscore}, which is given on the CARLA leaderboard website \cite{carlaleaderboard}.
        For the leaderboard to be a fair platform, it is crucial that \Cref{eq:dscore} is implemented correctly.
        This is hard to test using traditional techniques as we cannot reliably force particular infractions without incurring the considerable overhead of building a custom ego-vehicle and specially controlled driving environment.
        \begin{equation}\label{eq:dscore}
        \textit{InfractionPenalty}
        \times \textit{CompletionScore}
        \times (1 - \textit{OutsideLane})
        \end{equation}

        % This allows us to explore RQ\ref{rq:controlability:parameters} as we are not able to reliably control how or when infractions are committed.
        % \RH{Unclear? You expect the reader to remember what RQ1a is?}
        % Furthermore, as we will show in \Cref{sec:studied-cases}, there is also \emph{interaction} between variables, thereby allowing us to explore challenge \ref{ch:interaction} from \Cref{sec:introduction}.

    \item \label{req:ego-vehicle} (Ego-vehicle model).
        % \ref{req:ego-vehicle} concerns the generalisability of the ADSs between different models of vehicle.
        Human drivers are expected to adapt well to new models of vehicles, for example, when they buy a new car.
        It would be beneficial if ADSs could also achieve this, as it would mean they would not need to be retrained every time a new car was released.
        Our objective is to test whether the model of the ego-vehicle has a causal effect on the number of infractions that occur.
        This cannot be tested using traditional techniques as it is a statistical property over multiple runs.
        While SMT could test this property, it requires a large amount of highly controlled test data.

        % This allows us to explore RQ\ref{rq:controlability:data}. Here, we compare Causal Testing and SMT in terms of the data they require and the outcomes they produce.
        % The inspiration to test this property came during our data collection, when a few TCP scenarios from the privileged driver terminated suspiciously, reporting that an infraction had occurred before the scenario had even been fully initialised, but only for the default vehicle, the Lincoln MKZ2017. Our objective is to investigate whether the model of ego-vehicle has a causal effect on the infractions which occur and, if so, why.
        % \RH{I suggest some care here. An unfair reading of this would be: we spotted something interesting and so found a way of using our technique to find this thing (that we had already spotted anyway). }

    \item \label{req:carla-version} (CARLA version).
        This case explores a regression testing scenario between different CARLA versions.
        Our subject ADSs are designed to run on CARLA v0.9.10.1, but several versions of CARLA have subsequently been released.
        Since the changelog \cite{changelog} does not suggest that any changes or additional features should significantly affect performance, the ADSs should run equally well (if not better) on newer CARLA versions.
        Here, we test that newer versions of CARLA do not adversely impact the real-world time taken to simulate in-simulation time.
        As with \ref{req:ego-vehicle}, this is a statistical property over multiple runs.
        Previous approaches to Causal Testing  \cite{clark2022ctf} cannot test this, as there are unobservable variables at play.
        That is, the numbers of pedestrians and NPC vehicles are not recorded in the CARLA logs by default. They are unobserved.

        % This allows us to explore RQ\ref{rq:observability} and investigate how IV methods from CI can be used to circumvent the problem of unobservable variables, thereby addressing challenge \ref{ch:observable} from \Cref{sec:introduction}.
\end{enumerate}

We test our three requirements (the units of analysis) on four driving agents (which form sub-units of analysis) and analyse the resulting evidence separately.
We used the top two ADSs on the CARLA leaderboard with available and reusable code: TCP \cite{wu2022TCP} and CARLA Garage \cite{jaeger2023garage}.
Each has two kinds of driving agents (privileged and trained), which behave very differently.
\emph{Privileged agents} have access to ``privileged'' information such as the road layout and the locations of the other agents.
This makes them excellent drivers who commit very few infractions.
\emph{Trained agents} are machine learning models trained on data collected by a privileged agent.
They drive using only non-privileged data sources such as image data and LiDaR.
Thus, they typically commit more infractions than the privileged agents.
To make our study as diverse as possible, we consider one of each kind of agent for each ADS.

%% file: sections/04-studied-cases.tex
\section{Data Collection and Analysis}\label{sec:studied-cases}
This section presents how we obtained and analysed the evidence that we will use to answer our RQs in \Cref{sec:results}.
Our evidence is obtained by applying the \textit{four steps of causal testing} outlined in \Cref{sec:causal-testing}.
The first two steps (constructing the DAG and collecting test data) are shared between the three requirements.
The last two steps (defining and evaluating causal test cases) are unique to each requirement.
In particular, we consider how different estimation techniques, including interaction and IV methods, lead to different causal effect estimates and test outcomes.
Our replication package\footnote{https://github.com/CITCOM-project/carla-case-study} includes the data and code used to answer the RQs in this paper, as well as the artefacts (causal DAGs, test code and ADS setup) required to reproduce the results from fresh executions of CARLA.
% The test outcomes form part of the evidence we use to answer our RQs rather than being a result in themselves.

\subsection{Step 1: \CTstepDAG}\label{sec:cs-dag}
\begin{figure*}
  \centering
  \resizebox{0.8 \textwidth}{!}{\input{figures/dag}}
  \caption{The causal DAG for all three requirements, with the variables relevant to each requirement highlighted.
  The specific causal edges of interest are emboldened for clarity.
  Unobservable variables are drawn in grey. The dashed edge represents effect modification.
  We use the notation proposed in \cite{weinberg2007modification} of drawing (dashed) edges from nodes to other edges.}
  \Description{The causal DAG for all three requirements, with the variables relevant to each requirement highlighted.
  The specific causal edges of interest are emboldened for clarity.
  Unobservable variables are drawn in grey. The dashed edge represents effect modification.
  We use the notation proposed in \cite{weinberg2007modification} of drawing (dashed) edges from nodes to other edges.}
  \label{fig:dag}
\end{figure*}
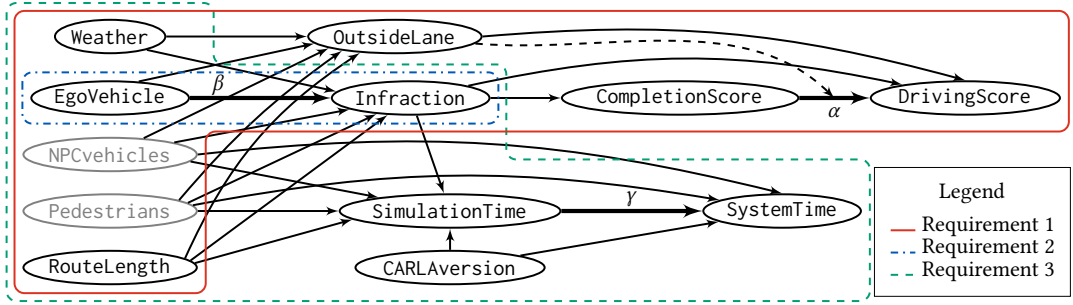

The first step of Causal Testing is to construct a causal DAG to represent the system.
This is shown in \Cref{fig:dag}, and is shared between our three requirements.
We used the CARLA documentation \cite{carlagithub,carlaleaderboard} and domain knowledge to determine which variables were relevant to our three requirements and how they related to each other.
The root nodes (\var{Weather}, \var{EgoVehicle}, \var{NPCvehicles}, \var{Pedestrians}, and \var{RouteLength}) represent CARLA configuration inputs.
The other nodes represent outputs, and will be discussed when we test the relevant requirements.
We used the method described in \cite{clark2022ctf} to determine the connections between the nodes.
Specifically, we assume that inputs are independent of each other (since they are chosen by the tester), and prune connections between the remaining nodes based on our knowledge of the system.

\subsection{Step 2: \CTstepData}\label{sec:cs-data}
The second step of Causal Testing is to collect test data.
To evaluate our requirements, we need each of our four driving agents to drive multiple models of ego-vehicle (\ref{req:ego-vehicle}) using multiple versions of CARLA (\ref{req:carla-version}).
We need to record their infractions (\ref{req:infraction-penalties} and \ref{req:ego-vehicle}) and the simulation runtimes (\ref{req:carla-version}).
While we can control the version of CARLA and the model of ego-vehicle, we cannot force infractions to occur, nor can we control the runtime of the simulation.
We must allow these to happen ``naturally'', as they normally would.
The advantage of Causal Testing is that the DAG in \Cref{fig:dag} allows us to identify and adjust for any bias this introduces \cite{clark2022ctf}.

To generate our test data, we followed the data collection instructions on the README page of each ADS, using the provided driving scenarios for the Town 01 map, which is the simplest of 12 road layouts supported by CARLA.
These scenarios are primarily intended for training and evaluating the respective ADSs.
TCP is distributed with 300 scenarios for Town 01.
CARLA Garage has just 132 scenarios, which are distinct from those of TCP.
We ran the scenarios for two versions of CARLA (v0.9.10.1 and v0.9.11) and two models of ego-vehicle (the default Lincoln MKZ2017 and the BMW Isetta), terminating execution at the first infraction so that a maximum of one infraction per scenario is considered.
%All our data and the scripts used to generate it are included in our replication package \cite{replication}.

Crucially, we did not generate the driving scenarios ourselves; we chose not to apply state-of-the-art ADS test scenario generation techniques here to maintain focus on the Causal Testing methodology as a whole.
This means our test data may not achieve the best coverage or produce the best test outcomes, but it makes it well suited to evaluate RQ\ref{rq:controlability:data} as it represents the kind of pre-existing runtime data with which Causal Testing is intended to be used \cite{clark2022ctf}.

To judge the accuracy and reliability of the IV methods we use to test \ref{req:carla-version}, we modified the code of TCP to enable the numbers of pedestrians and NPC vehicles (i.e., traffic participants) to be customised and recorded.
We then randomly spawned between 80 and 200 of each.
This was a non-trivial process that required several code files to be modified, but it allowed us to perform traditional adjustment, which we use as a ``gold standard'' to answer RQ\ref{rq:observability}.
It also enabled us to more effectively investigate RE1 as it yielded test data with more runs containing an infraction than the default value of 120.
We did not modify CARLA Garage except to change the model of the ego-vehicle, which is hardcoded.

% \begin{table}[ht!]
%   \centering
%   \begin{tabular}{lcccc}
%     \toprule
%      & \makecell{TCP\\Privileged} & \makecell{TCP\\Trained} & \makecell{Garage\\Privileged} & \makecell{Garage\\Trained}\\
%     \midrule
%     \multirow{2}{*}{\makecell{CARLA\\v0.9.10.1}} & 300 Lincoln & 300 Lincoln & 132 Lincoln & 132 Lincoln \\
%      & 300 BMW       & 300 BMW & 132 BMW     & 132 BMW \\
%     \midrule
%     \multirow{2}{*}{\makecell{CARLA\\v0.9.11}}   & 300 BMW   & 300 Lincoln & 132 BMW   & 132 Lincoln \\
%      & 300 Lincoln   & 300 BMW & 132 Lincoln & 132 BMW \\
%     \bottomrule
%   \end{tabular}
%   \caption{Number of data points collected for each ADS agent for each version of CARLA and each model of ego-vehicle.}
%   \label{tab:data-collection}
% \end{table}

% \footnotetext{As discussed in \Cref{sec:movivating-example}, the number of agents which successfully spawn may be less than this value if spawning a particular agent fails, for example because there is already an agent in the target location. This is not something which can be precisely controlled.}

% Infraction penalties
\subsection{RE1: Infraction Penalties}\label{sec:infraction-penalties}
Having drawn the DAG (step 1) and collected test data (step 2), we now define (step 3) and evaluate (step 4) causal test cases for \ref{req:infraction-penalties} to test that the correct penalty is applied for each infraction.

\subsubsection{Step 3: \CTstepCases} % Causal test case
\Cref{def:causal-test-case} states that a causal test case has three components: a treatment, an outcome, and an expected causal effect.
We want to test that the \emph{infraction penalty} is correct.
This corresponds to $\alpha$ in \Cref{fig:dag}, which is the effect of \var{CompletionScore} on \var{DrivingScore}.
Thus, these are our treatment and outcome, respectively.

There are four possible infractions in the Town 01 map: collisions with pedestrians, vehicles, and objects, and running red lights.
We define one causal test for each, and one for no infraction.
These all have \var{CompletionScore} as the treatment and \var{DrivingScore} as the outcome.
The expected causal effect is the corresponding infraction penalty, taken from the CARLA leaderboard \cite{carlaleaderboard} (see \Cref{tab:cs1-results}).
% We can be very precise with this because it is both known in advance and calculated deterministically.

\subsubsection{Step 4: \CTstepEvaluate} % Execution
We can now evaluate our five causal test cases.
As discussed in \Cref{sec:causal-testing}, this process has three substeps: identification, estimation, and comparison to the expected causal effect.

\textbf{Identification.}
The first step is to use \Cref{fig:dag} to identify sources of bias that must be adjusted to obtain an unbiased estimate.
\Cref{fig:dag} shows that \var{Infraction} is a common cause of \var{CompletionScore} and \var{DrivingScore}.
To adjust for this, the data are grouped into \emph{strata} by \var{Infraction}, and each test case is evaluated using the corresponding stratum.
Because some infractions happened more than others (see \Cref{tab:cs1-results}), the strata are not all the same size, which will affect the resulting estimate.
We discuss this further in \Cref{sec:results}.

The dashed edge in \Cref{fig:dag} shows that \var{OutsideLane} is an \emph{effect modifier} of \var{CompletionScore} on \var{DrivingScore}.
This stems from \Cref{eq:dscore}, which contains the term $\var{ComplectionScore} \times \var{OutsideLane}$, indicating that the two variables interact.
If we do not adjust for this, our estimates of the infraction penalties will be biased, potentially leading to unreliable test results.

\textbf{Estimation.}
Having identified the sources of bias, we now estimate the causal effect $\alpha$.
We use the regression model in \Cref{eq:alpha} for this, where $c$ is a constant.
This has the same form as \Cref{eq:dscore} from the CARLA leaderboard website \cite{carlaleaderboard}, except that \var{Infraction penalty} is replaced with $\alpha_1$, and a constant term has been added for completeness\footnote{If we were not adjusting for \var{Infraction} by stratification, it would also need a term in the equation.}.
If the \var{DrivingScore} is being calculated according to \Cref{eq:dscore}, the estimated coefficient $\alpha_1$ (i.e. the unit ATE) should equal the penalty for each infraction, as discussed in \Cref{sec:causal-testing}.
\begin{equation}
    \alpha_1 \times \var{CompletionScore} \times (1 - \var{OutsideLane}) + c
    \label{eq:alpha}
\end{equation}

Note that expanding out the bracket in \Cref{eq:dscore} gives the \emph{interaction term} $\var{CompletionScore} \times \var{OutsideLane}$.
This is how we adjust for the effect modification bias.
% Without this term, $\alpha_1$ would have to serve this role as well, making it a hybrid of both the infraction penalty and \var{OutsideLane}.
%
To investigate the importance of effect modification, we also consider \Cref{eq:alpha-simple}, which ignores the effect modification of \var{OutsideLane}.
This represents a naive estimation that only adjusts for confounding (by stratifying the data).
\Cref{fig:dag} indicates that this should produce a biased estimate.
\begin{equation}
  \alpha_2 \times \var{CompletionScore} + c
  \label{eq:alpha-simple}
\end{equation}

\textbf{Expected Effect.}
We determine the test outcomes by comparing our estimates for $\alpha_1$ and $\alpha_2$ to the expected causal effects, which are the infraction penalties from the CARLA leader board \cite{carlaleaderboard}.
\Cref{tab:cs1-results} shows these values, and is divided into two sections.
The first section shows the estimates for $\alpha_1$ in \Cref{eq:alpha}.
The second section shows the estimates for $\alpha_2$ in \Cref{eq:alpha-simple}.
% The third section shows how many times each ADS committed each infraction.
Missing entries correspond to infractions that were never committed.
Infractions that only occurred once did not give sufficient data to estimate confidence intervals.

\begin{table*}[!ht]
  \centering
  \small
    \caption{Test outcomes with estimated $\alpha_1$ as per \Cref{eq:alpha}, and $\alpha_2$ as per \Cref{eq:alpha-simple} to 3 decimal places. Failing test cases are highlighted with an (*) symbol. Missing values are shown with a (-) symbol.}
  \input{figures/case-1-results}

  \label{tab:cs1-results}
\end{table*}

% Neither of the privileged drivers collided with any pedestrians, and the CARLA Garage privileged driver did not crash into any layout objects (buildings, lamp posts, etc.).
% Because we are splitting the data into strata, by \var{Infraction} to adjust for its confounding, these strata will be empty, meaning that no estimate could be calculated.

\textbf{Test Outcomes.}
The top five rows of \Cref{tab:cs1-results} show that the estimates of $\alpha_1$ in \Cref{eq:alpha} are as expected.
The identical confidence intervals come from the infraction penalty being deterministic, so there is no variation in the dataset.
Thus, the regression model perfectly fits the data.
Every test case that we could evaluate passed for all four ADSs.

The bottom five rows of \Cref{tab:cs1-results} show that ignoring the effect modification from \var{OutsideLane} can lead to unreliable test outcomes.
For the privileged drivers, the results are unaffected because they never went \var{OutsideLane}, thereby nullifying the bias.
However, the estimates for the trained drivers are less precise than for $\alpha_1$, and two test cases fail because the effect estimates are not close enough (in this case, within 5\%) to the expected value.
This is because \Cref{eq:alpha-simple} does not include an interaction term, meaning $\alpha_2$ aggregates the infraction penalty and the proportion of the route spent \var{OutsideLane}.

% A further point of interest is that, while the TCP trained driver spent an average of 12.56\% of the route outside of its lane for routes where a layout collision was committed, the test case still passes.
% This is because there is a sufficiently small number of these routes that the distribution of time spent outside lane in each route has a large effect. In one particular case, over 90\% of a route was spent out of lane. This affects the \var{OutsideLane} average hugely, but has little effect on the estimate of $\alpha$ since this is just one data point and the infraction penalty for layout collision is fairly small.
% \todo[inline]{Not sure how relevant this actually is. If we drop out the percentage of time spent outside lane, we can probably get rid. Thoughts?}

\subsection{RE2: Ego-Vehicle Model}\label{sec:ego-vehicle}
Let us now define and evaluate causal test cases for \ref{req:ego-vehicle} from \Cref{sec:research-design}, which tests that the model of ego-vehicle does not have a significant effect on the infractions committed.

\subsubsection{Step 3: \CTstepCases} % Causal test case
As for \ref{req:infraction-penalties}, we first define the treatment, outcome, and expected causal effects.
We want to test that the model of \var{EgoVehicle} does not impact the \var{Infraction}s we observe.
Hence, the \var{EgoVehicle} is the treatment, and \var{Infraction} is the outcome.
Our expected causal effect ($\beta$ in \Cref{fig:dag}) is zero, since a good ADS should intuitively perform equally well on any vehicle, just as we would expect from a human driver.

\subsubsection{Step 4: \CTstepEvaluate} % Execution
Having defined our causal test case, we now carry out identification, estimation, and comparison to the expected causal effect.

\textbf{Identification.}
\Cref{fig:dag} shows that there is no bias that needs adjusting for here.
\var{EgoVehicle} is an input to the software, so there are no common causes or effect modifiers.

\textbf{Estimation.}
Since there are no sources of bias to adjust for, and we do not have a predefined equation to relate \var{EgoVehicle} and \var{Infraction}, we simply fit a model of the form $\var{Infraction} = \beta \times \var{EgoVehicle} + c$, where $\beta$ is the causal effect.
We here identify each \var{Infraction} by its numeric penalty rather than its name, as we did in \Cref{sec:infraction-penalties}.
Since our test data includes every driving scenario run with both models of the ego-vehicle, Causal Testing using the full dataset effectively becomes SMT.
To help answer RQ\ref{rq:controlability:data}, which concerns uncontrolled data, we additionally estimate $\beta$, using the first half of the driving scenarios for the Lincoln, and the second half for the BMW, for each version of CARLA.
When the data is partitioned in this way, no route is driven by both ego-vehicles, so SMT is not directly applicable.
However, we expect Causal Testing to produce similar estimates that lead to the same test outcomes.

\textbf{Expected Effect.}
To determine the test outcomes, we compare our estimates of $\beta$ to the expected causal effect (i.e., zero).
As mentioned in \Cref{sec:causal-testing:eval}, the absence of a causal effect is indicated by the confidence intervals for the estimate containing zero.
\Cref{tab:cs2-results} shows our estimates for each of the four drivers.
The second column shows our estimates calculated using the full dataset, where each agent drove both vehicles for all scenarios.
The last column shows our estimates calculated using the dataset, where each agent drove each vehicle for half of the scenarios.
As expected, the two columns show very similar causal effects for each driver.

% Note that, although we do not know the true values of $\beta$, we can treat the estimates calculated by the SMT style as the ``gold standard'' since it is the most accurate estimate we can calculate with the data we have. \DS{To me, this part seems a bit unnecessary...}

% Recall from \Cref{sec:causal-testing} that is indicated by the confidence intervals for the causal effect estimate containing zero.
% \Cref{tab:cs2-results} shows these estimates for each of the four drivers.
% The first column shows our estimates of $\beta$ calculated using the full dataset, in the SMT setting where each driver drove each vehicle for every scenario.
% Since we do not know for sure the true values of $\beta$, we here treat this as the ``gold standard'' as it is the most accurate estimate we can calculate with the data we have.
% The second column shows our estimates of $\beta$ using the dataset where each driver only drove each vehicle for half of the scenarios.
% As we expected, the two columns show very similar causal effects for each driver.

\begin{table}[t]
  \centering
  \small
    \caption{Estimated effect on the infraction penalty of changing the model of ego-vehicle from the Lincoln MKZ2017 to the BMW Isetta. Failing test cases are highlighted with (*).}
  \input{figures/case-2-results}

  \label{tab:cs2-results}
\end{table}

\textbf{Test Outcomes.}
\Cref{tab:cs2-results} shows that the test case for the CARLA Garage privileged driver passes. The confidence intervals contain zero, indicating no significant causal effect.
The other three drivers fail.
Both trained drivers have a negative effect of around $-0.1$ with confidence intervals that do not contain zero.
This indicates that the BMW leads to worse driving than the Lincoln. This is not surprising as the drivers were only trained in the Lincoln, but the result is still cause for concern and is discussed further in \Cref{sec:results}.

More surprisingly, the TCP privileged driver seems to drive \emph{better} in the BMW than in the Lincoln.
While the effect size is very small, the confidence intervals do not contain zero, so the test case fails.
This is surprising, and we will discuss the underlying causes and implications of this in \Cref{sec:results}.

\subsection{RE3: CARLA Version}\label{sec:carla-version}
We now define and evaluate the causal test case for \ref{req:carla-version}.
This considers a regression testing scenario to validate that updating CARLA from v0.9.10.1 to v0.9.11 does not adversely affect the performance.

\subsubsection{\CTstepCases} % Causal test case
% Again, we must define the treatment, outcome, and expected causal effect.
Our expected causal effect is ``not positive'', as we do not anticipate a slower simulation, but we do not mind if it speeds it up or stays the same.

Defining the treatment and outcome is a little more complex than the first two requirements.
Simply testing the causal effect of the \var{CARLAversion} on the \var{SystemTime} does not incorporate the actual performance of the simulation, i.e. how much real-world time it takes to simulate each second of in-simulation time.
This is characterised by the direct causal effect of \var{SimulationTime} on \var{SystemTime}.
We need to test that this causal effect stays the same \emph{between} the versions of CARLA.
Thus, \var{SimulationTime} is our treatment and \var{SystemTime} is our outcome.

\subsubsection{\CTstepEvaluate} % Execution
Having defined our causal test case, we now carry out identification, estimation, and comparison to the expected causal effect.

\textbf{Identification.}
\Cref{fig:dag} shows that there are three sources of bias here: the \var{CARLAversion}, and the numbers of \var{NPCvehicles} and \var{Pedestrians}, as these are all common causes of our treatment and outcome.
The intuition for this is that heavy traffic may lead to routes taking more simulation time to complete and more real-world time per time step, as there are more agents to update.

Unlike \ref{req:ego-vehicle}, SMT is not applicable here since, as mentioned in \Cref{sec:movivating-example,sec:cs-data}, we cannot precisely control the values of \var{NPCvehicles} and \var{Pedestrians}.
Indeed, we are not even able to \emph{observe} these values by default, as they are not logged by TCP or CARLA Garage.
This means that we cannot collect the controlled data necessary to perform SMT.

\textbf{Estimation.}
Having established our sources of bias, we can now estimate the causal effects.
Since we are comparing causal effects between versions of CARLA, we adjust for the bias from \var{CARLAversion} by considering the two versions separately.
The adjustment for \var{NPCvehicles} and \var{Pedestrians} is more nuanced, and we will consider and compare three different estimation methods here in order to obtain sufficient evidence to answer RQ\ref{rq:observability}.

Firstly, because the values of \var{NPCvehicles} and \var{Pedestrians} are not logged by default, we will use IV methods (see \Cref{sec:instrumental-variables}) to estimate our causal effect $\gamma$ without needing this data.
Using \var{RouteLength} as the \emph{instrument}, we can estimate $\gamma$ by dividing its total causal effect on \var{SystemTime} (\Cref{eq:length-sys}) by its direct effect on \var{SimulationTime} (\Cref{eq:length-sim}).
\begin{subequations}\label{eq:iv}%
	\begin{align}%
		\var{SystemTime} &= \gamma_{sys}\times\var{RouteLength}   \label{eq:length-sys} \\
		\var{SimulationTime} &= \gamma_{sim}\times\var{RouteLength}   \label{eq:length-sim} \\
        \gamma &= \nicefrac{\gamma_{sys}}{\gamma_{sim}}
	\end{align}%
\end{subequations}

We use \var{RouteLength} as the instrument because it matches the causal structure in \Cref{fig:iv}.
That is, there is no edge between \var{Pedestrians} or \var{NPCvehicles} and \var{RouteLength}, and there is a path $\var{RouteLength} \to \var{SimulationTime} \to \var{SystemTime}$.
While we cannot know for sure that the relationships are linear, the intuition is that the longer a route is, the more simulation time it should take, since the ego-vehicle has to travel further, so more wall-clock \var{SystemTime} is required to run the simulation.
% Even if the relationships are not precisely linear, our test oracle (the expected causal effect) is fairly weak --- we only seek to establish that the causal effect is not positive --- so our IV estimates should be able to establish this as long as the relationships are at least monotonic, which the above intuition supports.

Secondly, since we have no way of knowing the true causal effect, we created an artificial ``gold standard'' (as discussed in \Cref{sec:cs-data}) by modifying TCP to record the numbers of pedestrians and NPC vehicles to enable traditional adjustment \cite{pearl2009causality} using \Cref{eq:gamma} to compute estimates that are as accurate as possible.
We gained the same data for CARLA Garage by manual code inspection, revealing that it always spawns 120 NPC vehicles and either zero or one pedestrian, depending on the driving scenario.
In both cases, the information was time-consuming to obtain, and may not be obtainable at all in the general case.
\begin{equation}
\begin{aligned}
  \var{SystemTime} =\, \gamma\times\var{SimulationTime}\,+ \\
                       c_1\times\var{Pedestrians}\,+ c_2\times\var{NPCvehicles}+ c_3
  \label{eq:gamma}
\end{aligned}
\end{equation}

Finally, we consider \Cref{eq:gamma-simple}, which ignores the confounding effect of \var{Pedestrians} and \var{NPCvehicles}.
\Cref{fig:dag} indicates that this should give a biased estimate.
This may be more \emph{accurate} (i.e. closer to traditional adjustment) than IV methods, if the bias is sufficiently weak.
However, this cannot be determined in advance.
\begin{equation}
   \var{SystemTime} = \gamma\times\var{SimulationTime} + c
   \label{eq:gamma-simple}
\end{equation}

\textbf{Expected Effect.}
We now determine the test outcomes by comparing estimates of $\gamma$ between CARLA versions.
\Cref{tab:cs3-results} shows the estimated direct causal effect of \var{SimulationTime} on \var{SystemTime}, i.e. how long it takes to simulate one second of time in simulation ($\gamma$ in \Cref{fig:dag}).
We here expect the causal effect to be zero.
% The first part shows the estimates calculated using IV methods, as described by \Cref{eq:iv}.
% The second part shows the same estimates calculated using traditional adjustment (\Cref{eq:gamma}), which we use as a ``gold standard'' for the value of $\gamma$.
% The final part shows the estimates calculated without adjustment \Cref{eq:gamma-simple}.

\begin{table}[t]
  \centering
  \small
  \caption{Estimated direct causal effect of simulation time on system time for the different versions of CARLA. This represents how much real-world time it takes to simulate one second of the simulation.}
  \input{figures/case-3-results}
  \label{tab:cs3-results}
\end{table}

\textbf{Test Outcomes.}
In ~\Cref{tab:cs3-results}, the first four rows show the estimates for $\gamma$ calculated using IV methods.
This shows that CARLA 0.9.11 is slower for all four drivers.
The confidence intervals for the corresponding estimates between different versions of CARLA do not overlap, so the test cases all fail.

The second four rows show that classical adjustment gives similar estimates, although the confidence intervals between CARLA versions for the CARLA Garage privileged driver overlap very slightly, leading to a passing test result.
For the other drivers, there is no overlap and test cases still fail.

The final four rows show that the biased estimates calculated without adjustment are very close to those calculated with classical adjustment and produce the same test outcomes.
We will discuss this further in \Cref{sec:results:observability}

%% file: figures/dag.tex
\begin{tikzpicture}[node distance=0.3cm and 2cm,->,> = latex']
  \node[vertex] (Weather) {\var{Weather}};
  \node[vertex] (Ego-vehicle)      [below=of Weather, yshift=-2]{\var{EgoVehicle}};
  \node[vertex, gray] (NPC vehicles)     [below=of Ego-vehicle] {\var{NPCvehicles}};
  \node[vertex, gray] (Pedestrians)      [below=of NPC vehicles] {\var{Pedestrians}};
  \node[vertex] (Route length)     [below=of Pedestrians] {\var{RouteLength}};
  \node[vertex] (Infraction)       [right=of Ego-vehicle] {\var{Infraction}};
  \node[vertex] (Outside lane)       [right=of Weather] {\var{OutsideLane}};
  \node[vertex] (Completion score) [right=of Infraction, shift={(-1, 0)}] {\var{CompletionScore}};
  \node[vertex] (Driving score)    [right=of Completion score, shift={(-1, 0)}] {\var{DrivingScore}};
  \node[vertex] (Simulation time)  [right=of Pedestrians] {\var{SimulationTime}};
  \node[vertex] (System time)      [right=of Simulation time] {\var{SystemTime}};
  \node[vertex] (CARLA version)          [below=of Simulation time] {\var{CARLAversion}};

  \draw[edge, line width=0.6mm] (Completion score) to node (alpha) [below] {$\alpha$} (Driving score);
  \draw[edge] (CARLA version) to (Simulation time);
  \draw[edge] (CARLA version) to (System time);
  \draw[edge] (Route length.3) to (Simulation time.185);
  \draw[edge] (Route length.5) to (Infraction);
  \draw[edge] [bend left=16] (Route length.7) to (Outside lane.205);
  \draw[edge, line width=0.6mm] (Simulation time) to node (gamma) [above, shift={(0, -0.07)}] {$\gamma$} (System time);
  \draw[edge] (NPC vehicles.-5) to (Simulation time);
  \draw[edge] [bend left=10] (NPC vehicles.0) to (System time.90);
  \draw[edge] (NPC vehicles) to (Infraction);
  \draw[edge] (NPC vehicles) to (Outside lane.190);
  \draw[edge, line width=0.6mm] (Ego-vehicle) to node (beta) [above, pos=0.2, shift={(0, -0.1)}] {$\beta$} (Infraction);
  \draw[edge] (Ego-vehicle.30) to (Outside lane.185);
  \draw[edge] (Weather.-20) to (Infraction.175);
  \draw[edge] (Weather) to (Outside lane.180);
  \draw[edge] (Infraction) to (Completion score);
  \draw[edge] [bend left=12] (Infraction) to (Driving score);
  \draw[edge] [bend left=10] (Outside lane.east) to (Driving score.north);
  \draw[edge] (Infraction) to (Simulation time);
  \draw[edge] (Pedestrians) to (Simulation time);
  \draw[edge] [bend left=10] (Pedestrians.4) to node (key) {} (System time.170);
  \draw[edge] (Pedestrians.6) to (Infraction);
  \draw[edge] [bend left=7] (Pedestrians.9) to (Outside lane.195);

  \draw[edge, dashed, looseness=0.9, out=-5] (Outside lane) to ($(alpha)+(0, 0.2)$);
  % \draw[edge, dashed, out=30, in=-120, looseness=0.5] (CARLA version) to ($(gamma)+(0, -0.25)$);

  \node[draw=none, rectangle] (cs2) [fit=(Ego-vehicle) (Infraction)] {};

  \node[draw=none, rectangle] (cs1 inputs) [fit=(Weather) (Ego-vehicle) (NPC vehicles) (Pedestrians) (Route length)] {};
  \node[draw=none, rectangle] (cs1 outputs) [fit=(cs2.south) (Outside lane) (Infraction) (Completion score) (Driving score)] {};

  \node[draw=none, rectangle] (cs3 left) [fit=(cs1 inputs)] {};
  \node[draw=none, rectangle] (cs3 mid) [fit=(cs2)] {};
  \node[draw=none, rectangle] (cs3 bot) [fit=(cs1 inputs.south) (gamma) (key) (Simulation time) (CARLA version) (System time)] {};

  % \node[draw=none, rectangle] (cs1c) [fit=(Infraction) (Outside lane) (Driving score.east |- Weather.north) (cs2.south)] {};
  % \node[draw=none, rectangle] (cs1b) [fit=(cs1c)] {};

  % \node[draw=none, rectangle] (cs3a) [fit=(cs1) (Route length)] {};
  % \node[draw=none, rectangle] (cs3b) [fit=(System time) (gamma) (Route length)] {};
  % \node[draw=none, rectangle] (cs3c) [fit=(cs2b.east |- cs1b.north) (cs2b) (Route length)] {};

  \draw[-,BLUE,rounded corners,thick,dashdotted] %
  (cs2.north west) -- %
  (cs2.south west) -- %
  (cs2.south east) -- %
  (cs2.north east) -- %
  cycle;

  \draw[-,GREEN,rounded corners,thick,dashed] %
  (cs3 left.north west) -- %
  (cs3 left.south west) -- %
  (cs3 bot.south east) -- %
  (cs3 bot.north east) -- %
  (cs3 mid.east |- cs3 bot.north) -- %
  (cs3 mid.north east) -- %
  (cs3 left.east |- cs3 mid.north) -- %
  (cs3 left.north east) -- %
  cycle;

  \draw[-,RED,rounded corners,thick] %
  (cs1 inputs.north west) -- %
  (cs1 inputs.south west) -- %
  (cs1 inputs.south east) -- %
  (cs1 inputs.east |- cs1 outputs.south) -- %
  (cs1 outputs.south east) -- %
  (cs1 outputs.north east) -- %
  cycle;

  % legend
  \node[draw=none, shift={(0.1, -0.05)}] (cs3) at (Driving score |- CARLA version) {\tikz{\draw[-,GREEN,thick,dashed]  (0,0)--(0.35,0);} Requirement 3};
  \node[draw=none] (cs2) at ($(cs3)+(0, 1em)$) {\tikz{\draw[-,BLUE,thick,dashdotted]  (0,0)--(0.35,0);} Requirement 2};
  \node[draw=none] (cs1) at ($(cs2)+(0, 1em)$) {\tikz{\draw[-,RED,thick]  (0,0)--(0.35,0);} Requirement 1};
  \node[draw=none] (legend label) at ($(cs1)+(0, 1.5em)$) {Legend};
  \node[draw,shape=rectangle] (legend) [fit=(legend label) (cs3)] {};
\end{tikzpicture}

%% file: figures/case-1-results.tex
\newcommand{\emptyCI}{[\hspace{1.6ex}-\hspace{1.6ex}, \hspace{1.6ex}-\hspace{1.6ex}]}
\newcommand{\expected}{$\alpha_\text{Expected}$}
\newcommand{\eqa}{\multirow{5}{*}{\rotatebox[origin=c]{90}{\parbox{5em}{\centering \Cref{eq:alpha}\\$\alpha_1$}}}}
\newcommand{\eqb}{\multirow{5}{*}{\rotatebox[origin=c]{90}{\parbox{5em}{\centering \Cref{eq:alpha-simple}\\$\alpha_2$}}}}
\newcommand{\nos}{\multirow{6}{*}{\rotatebox[origin=c]{90}{\parbox{5em}{\centering Number of occurrences}}}}
\newcommand{\red}[1]{{*\hfill#1}}

\begin{tabular}{llrrrrr}
  \toprule
       & Infraction            & \expected{} & TCP Privileged      & TCP Trained         & CARLA Garage Privileged & CARLA Garage Trained      \\
  \midrule
  \eqa & No infraction         & 1.00        & 1.000[1.000, 1.000] & 1.000[1.000, 1.000] & 1.000[1.000, 1.000]     & 1.000[1.000, 1.000]       \\
       & Red light             & 0.70        & 0.700[0.700, 0.700] & 0.700[0.700, 0.700] & 0.700[-, -]             & 0.700[0.700, 0.700]       \\
       & Collisions layout     & 0.65        & 0.650[0.650, 0.650] & 0.650[0.650, 0.650] & -                       & 0.650[0.650, 0.650]       \\
       & Collisions vehicle    & 0.60        & 0.600[0.600, 0.600] & 0.600[0.600, 0.600] & 0.600[0.600, 0.600]     & 0.600[0.600, 0.600]       \\
       & Collisions pedestrian & 0.50        & -                   & 0.500[0.500, 0.500] & -                       & 0.500[-, -]               \\\midrule
  \eqb & No infraction         & 1.00        & 1.000[1.000, 1.000] & 1.024[1.005, 1.043] & 1.000[1.000, 1.000]     & 1.046[0.964, 1.128]       \\
       & Red light             & 0.70        & 0.700[0.700, 0.700] & 0.698[0.696, 0.700] & 0.700[-, -]             & 0.700[0.700, 0.700]       \\
       & Collisions layout     & 0.65        & 0.650[0.650, 0.650] & 0.650[0.620, 0.680] & -                       & \red{0.538[0.475, 0.601]} \\
       & Collisions vehicle    & 0.60        & 0.600[0.600, 0.600] & 0.604[0.594, 0.614] & 0.600[0.600, 0.600]     & \red{0.482[0.372, 0.591]} \\
       & Collisions pedestrian & 0.50        & -                   & 0.500[0.500, 0.500] & -                       & 0.500[-, -]               \\
  % \nos & No infraction         &             & 1122                & 879                 & 518                     & 461                       \\
  %      & Red light             &             & 40                  & 119                 & 1                       & 5                         \\
  %      & Layout collision      &             & 4                   & 47                  & 0                       & 45                        \\
  %      & Vehicle collision     &             & 34                  & 149                 & 9                       & 16                        \\
  %      & Pedestrian collision  &             & 0                   & 6                   & 0                       & 1                         \\
  %      & Total \% outside lane &             & 0                   & 1.978               & 0                       & 1.949                     \\
  \bottomrule
\end{tabular}

%% file: figures/case-2-results.tex
\setlength\tabcolsep{1mm}
\newcommand{\red}[1]{{*\hfill#1}}
\begin{tabular}{lrr}
  \toprule
  ADS                     & $\beta$ estimate       & $\beta$ estimate 1/2   \\\midrule
  TCP trained             & \red{-0.111[-0.131, -0.092]} & \red{-0.115[-0.143, -0.087]} \\
  TCP privileged          & \red{0.0132[0.003,0.023]}    & \red{0.018[0.005, 0.031]}    \\
  CARLA G. trained    & \red{-0.112[-0.135, -0.089]} & \red{-0.102[-0.133, -0.07]}  \\
  CARLA G. privileged & 0.006[-0.003,0.015]    & 0.003[-0.01, 0.016]    \\
  \bottomrule
\end{tabular}

%% file: figures/case-3-results.tex
\setlength\tabcolsep{1mm}
\newcommand{\gray}[1]{{\color{gray}#1}}
\newcommand{\instrumental}{\multirow{4}{*}{{\parbox{1.5cm}{\centering IV\\methods\\(\Cref{eq:iv})}}}}
\newcommand{\classical}{\multirow{4}{*}{{\parbox{1.5cm}{\color{gray} \centering Gold standard\\(\Cref{eq:gamma})}}}}
\newcommand{\none}{\multirow{4}{*}{{\parbox{1.5cm}{\centering No\\adjustment\\(\Cref{eq:gamma-simple})}}}}
  \begin{tabular}{llll}
    \toprule
                  & Driver                  & CARLA v0.9.10.1      & CARLA v0.9.11        \\
    \midrule
    \instrumental & TCP trained             & 4.470[4.401, 4.569] & 6.829[6.766, 6.886] \\
                  & TCP privileged          & 3.886[3.837, 3.938] & 6.306[6.253, 6.364] \\
                  & Garage trained    & 6.412[6.199, 6.595] & 8.522[8.312, 8.767] \\
                  & Garage Privileged & 7.751[7.389, 8.100] & 8.383[8.114, 8.617] \\
    \midrule
    \classical    & \gray{TCP trained      } & \gray{4.523[4.437, 4.609]} & \gray{6.677[6.605, 6.749]} \\
                  & \gray{TCP privileged   } & \gray{3.838[3.759, 3.918]} & \gray{6.180[6.101, 6.259]} \\
                  & \gray{Garage trained   } & \gray{6.779[6.696, 6.861]} & \gray{9.009[8.875, 9.143]} \\
                  & \gray{Garage Privileged} & \gray{7.162[6.883, 7.441]} & \gray{7.814[7.398, 8.231]} \\
    \midrule
    \none         & TCP trained             & 4.522[4.437, 4.607] & 6.682[6.611, 6.753] \\
                  & TCP privileged          & 3.832[3.755, 3.909] & 6.182[6.104, 6.260] \\
                  & Garage trained    & 6.779[6.696, 6.861] & 9.009[8.875, 9.143] \\
                  & Garage Privileged & 7.162[6.883, 7.441] & 7.814[7.398, 8.231] \\
    \bottomrule
\end{tabular}

%% file: sections/05-results.tex
\section{Analysis and Answers to Research Questions}\label{sec:results}
This section answers our RQs using the evidence from \Cref{sec:studied-cases}.
% For each question, we first present generalized findings, and then discuss the evidence upon which these findings are based in the context of our three tested requirements.
% To answer RQ\ref{rq:controlability}, we consider the effect estimates and test outcomes for \ref{req:infraction-penalties} and \ref{req:ego-vehicle}.
% To answer RQ\ref{rq:observability}, we consider the effect estimates and test outcomes for \ref{req:carla-version}.

\subsection{RQ\ref{rq:controlability:parameters} Can Causal Testing deliver reliable test outcomes for software with interacting parameters?}
% R1
In \ref{req:infraction-penalties}, the infraction penalty is the direct causal effect of the \var{CompletionScore} on the \var{DrivingScore}.
The proportion of the route spent \var{OutsideLane} interacts with the \var{CompletionScore}, introducing a source of bias.
We used an \emph{interaction term} in the estimator to adjust for this, allowing us to validate the penalties for each infraction.
% When we used the same test data to estimate the causal effects without proper adjustment, the causal effect estimates were no longer exact, even though infraction penalties are calculated deterministically, so there should not be any variance in the data.
When we estimated the causal effects without adjusting for this bias, two test cases for the CARLA Garage trained driver failed (even though they should have passed) because the causal effect estimates were not within 5\% of the expected effect.
% Our main findings can be summarized as follows.
\begin{tcolorbox}[left=0.5mm, right=0.5mm, top=0.5mm, bottom=0.5mm]
    \emph{Interaction between parameters can cause tests to fail when there is no fault.
    Causal Testing enables us to isolate direct causal effects, even when variables interact.}
\end{tcolorbox}
% The nature of the causal relationships between variables can be affected by the values of other variables.
% If is not properly adjusted for, the resulting bias can cause test cases to fail as the estimated causal effect ends up as an aggregate of the causal effect and its modifier.

% \emph{Causal Testing enables us to isolate direct causal effects, even when variables interact.}
% Including an \emph{interaction term} in the causal effect estimator allows us to adjust for interaction between variables to isolate the direct causal effect.
% This gives us reliable test outcomes even when we cannot control variable values, and is particularly advantageous when testing the relationships between intermediate software outputs.

\subsection{RQ\ref{rq:controlability:data} Can Causal Testing deliver reliable test outcomes when using uncontrolled data?}

In \ref{req:infraction-penalties}, we stratified our test data by \var{Infraction} to adjust for its bias.
To achieve a similar result using SMT, we would need to control which infraction the ego-vehicle committed each test run.
This is impossible here due to the non-controllability of CARLA and the ADSs under test, meaning that some infractions did not occur often enough in the test data for us to calculate confidence intervals.
This highlights an inherent property of using uncontrolled data: we can only test aspects of behaviour that are covered by the data.

In \ref{req:ego-vehicle}, we discovered an unexpected direct causal link between the model of ego-vehicle and the infraction penalty for three of the four driving agents we tested (i.e. the choice of ego-vehicle model has a causal effect on the infractions committed).
Here, causal testing essentially reduces to SMT because there is no confounding between the treatment and outcome.
However, without the DAG in \Cref{fig:dag}, we would have no way to confirm this, so would not have been able to draw a causal conclusion.
We also investigated the ability of Causal Testing to obtain reliable test outcomes from uncontrolled data by evaluating the same test cases using just half of our test data, where no route was driven by both models of the ego-vehicle.
Where SMT is not directly applicable to this data, the causal tests all led to the same test outcomes as the full dataset.

We used the same test data for all three of our requirements.
We also used the same test scripts to validate four separate driving agents.
While it was computationally expensive to collect test data from each driver, the effort required to test all four systems was no more than testing just one.
% To summarize, our main findings are as follows.
\begin{tcolorbox}[left=0.5mm, right=0.5mm, top=0.5mm, bottom=0.5mm]
    \emph{Observing sufficiently many inputs is critical to estimate causal effects. However
    Causal Testing allows us to use less (and less controlled) test data than would be needed for SMT.
    It also enables us to reuse the same test data to test multiple requirements.}
\end{tcolorbox}

\subsection{RQ\ref{rq:observability}  Can Causal Testing support the testing of software with unobservable variables?}\label{sec:results:observability}

In \ref{req:carla-version}, we investigated the simulation performance of two versions of CARLA.
As with \ref{req:infraction-penalties}, SMT cannot be applied here as the causal effect of \var{SimulationTime} on \var{SystemTime} is confounded by the numbers of \var{Pedestrians} and \var{NPCvehicles}, which are not recorded in the logs by default.
Since we have no way of knowing the true causal effect, we recorded their values to enable us to use traditional adjustment as an artificial ``gold standard'' to compare estimates calculated using IV methods and without any form of adjustment.
IV methods gave a median error of 0.31, and no adjustment gave 0.001.
Since our estimates represent how many real-world seconds it takes to simulate one second of simulation time, these errors are barely perceptible.

While the estimates produced without adjustment are closer to the gold standard, the IV estimates still produce reliable test outcomes.
In the general case, where we could not obtain the values of confounding variables, we would have no way of knowing which estimate is more accurate.
However, IV methods produce causally valid and sufficiently accurate estimates as they adjust for the bias without needing to know the values of the confounders.
% To summarize, our main finding is as follows.
\begin{tcolorbox}[left=0.5mm, right=0.5mm, top=0.5mm, bottom=0.5mm]
    \textit{Causal Testing enables unbiased causal effect estimates to be calculated when we cannot observe certain variables, as long as certain assumptions are satisfied.}
\end{tcolorbox}

% While estimation without any adjustment led to more accurate estimates in this respect than IV methods, there is no way to determine this in advance.

% Where we have the causal structure shown in \Cref{fig:iv}, and can assume linear relationships between variables, IV methods can be used to adjust for the bias of confounding variables (i.e. common causes), although the estimates may not be as accurate as traditional adjustment.
% Where the bias is weak, estimation without adjustment may also produce more accurate values.
% However, there is no way to know this without being able to compare to traditional adjustment, which cannot be done if we do not have data for all the variables we need to adjust for.

\subsection{RQ\ref{rq:faults} Can Causal Testing  reveal faults in software with interacting and unobservable parameters?}\label{sec:results:faults}

While testing \ref{req:ego-vehicle}, we discovered two inconsistencies.
Firstly, the trained agents performed worse when driving ego-vehicles that they were not trained on.
This is concerning, as it suggests that expensive training data needs to be collected for every new vehicle.

Secondly, we discovered that the TCP privileged agent performs slightly better when driving the BMW rather than the Lincoln, which is unexpected as the privileged agent is not a trained model, so it should drive all vehicles equally well.
Inspecting the driving scenarios revealed that the ego-vehicle is sometimes spawned already committing an infraction (e.g. just in front of a red light).
The BMW experiences this less because it is smaller than the Lincoln.
While this behaviour is unexpected, we do not call it a ``bug'', as neither CARLA nor TCP is doing anything wrong.
It is just that some of TCP's driving scenarios represent unrealistic behaviour.

In \ref{req:carla-version}, we discovered a significant decrease in performance between CARLA v0.9.10.1 and CARLA v0.9.11.
Although the cause of this is beyond the scope of this paper, it suggests either a regression or an omission from the changelog \cite{changelog}.
Without Causal Testing, we would not be able to obtain reliable test outcomes for this requirement because we cannot control (or even observe, by default) the numbers of pedestrians and NPC vehicles within the simulation.
We would, therefore, only be able to conclude that the CARLA version was \emph{associated} with a change in runtime.
\begin{tcolorbox}[left=0.5mm, right=0.5mm, top=0.5mm, bottom=0.5mm]
    \textit{Causal Testing can discover faults in software with interacting and unobservable parameters.}
\end{tcolorbox}

%% file: sections/07-validity-threats.tex
\section{Threats to Validity}\label{sec:validity-threats}
\textit{External validity:}
We carried out an evaluative case study by testing three requirements surrounding ADS testing.
We chose this setting because it addresses the challenges posed in previous work on Causal Testing \cite{clark2022ctf,giamattei2024causality} (see \Cref{sec:movivating-example}).
As discussed in Section~\ref{sec:causal-testing}, Causal Testing can, in principle, test any behaviour framed as the effect of treatment on an outcome.
While underlying CI has been shown to be generally applicable \cite{bareinboim2016causal}, including for investigating complex non-linear relationships between variables, a broader study is required to establish the circumstances under which Causal Testing is applicable in the general case.

\textit{Internal validity:}
We selected our three requirements for their relevance to our research questions and drew the associated DAG ourselves.
This leads to the risk that the success of the approach has been biased: that our chosen requirements favour the technique.
This is an intrinsic risk to any case study and is integral to our future work.
However, it is worth noting that we did not control the test data; the driving scenarios formed part of the training data for TCP and Carla Garage.
The fact that the same routes could be used to address all three test objectives is a testament to a core attribute of CI (and Causal Testing) -- the fact that the approach used to analyse a test set is independent of the data.

%% file: sections/08-related-work.tex
\section{Related Work}\label{sec:related-work}
\textbf{ADS Testing.}
There is a wealth of literature on ADS testing \cite{zhong2021survey,sun2022scenario,tang2023survey,zhang2023finding}.
A key research topic in this area is the generation of driving scenarios that lead to misbehaviour \cite{haq2022efficient,deephyperion-cs}, with several recent techniques \cite{giamattei2024causality,jiang2024generation} employing causal reasoning.
Our RQ\ref{rq:controlability:data} showed Causal Testing is complementary to such approaches, as the generated scenario data can be used to explain \emph{why} particular scenarios failed, and reused to test additional causal relationships.
Along these lines, \citet{han2020fuzz} use metamorphic tests to answer questions like ``Would the ego-vehicle still have crashed into an object if it had been further away?''.
While such questions are clearly causal, they are answered via the controlled collection of new data, where our approach can use existing data.

\textbf{Machine Learning-Inferred Models of Tested Behaviour.}
In this work, we used causality-informed linear regression models to estimate causal effects.
This relates to a significant body of work on machine learning approaches for inferring models from test executions.
Such approaches often use off-the-shelf algorithms, such as linear regression \cite{arrieta2021using}, support vector regression \cite{chen2020learning}, and ensemble models \cite{haq2022efficient}.
Machine learning approaches have also been applied to ADSs to estimate the probability of safety violations \cite{nitsche2018junctions,norden2019estimation}.
A key limitation of these approaches is that the challenges outlined in \Cref{sec:movivating-example} --- namely, nondeterminism, observability, interaction, and long execution times --- typically prevent us from collecting a sufficiently large and diverse set of executions to characterise the underlying behaviour.
\citet{norden2019estimation} tackle this by limiting execution time, but this is not always feasible.

\textbf{Causality in Software Engineering.}
Causal reasoning is increasingly being applied in a range of software engineering contexts \cite{siebert2023applications,giamattei2025survey}.
The technique of Causal Testing was originally published in \citet{clark2022ctf}, with subsequent papers proposing techniques to automatically generate metamorphic test cases from causal DAGs \cite{clark2023metamorphic} and measure test adequacy \cite{foster2024adequacy}.

Causal reasoning is also popular in the field of fault localisation.
For example, \citet{johnson2020causal} explain the root cause of faulty software behaviour by mutating existing tests to form a suite of minimally different tests that are not fault-causing.
The test suites are then compared to understand \emph{why} a fault occurred.
Several techniques also employ CI, using the program dependence graph as a DAG \cite{baah2010causal,baah2011mitigating,podgurski2020counterfault,shu2013mfl, bai2015numfl, gore2012reducing}.

The great advantage of Causal Inference is the fact that it can be applied to observational data, without the need for a controlled experiment. In this context, it has also been shown to be a valuable tool for empirical software engineering. Recent work by \citet{furia2023towards} has shown how it can be more precise at analysing programmer performance than purely predictive techniques.

\textbf{Automatic Generation of DAGs.}
While manual creation of DAGs is widely accepted in fields such as epidemiology, causal discovery \cite{malinsky2018causal} aims to automatically learn causal structures from data by exploiting asymmetries that separate association from causation \cite{glymour2019review}.
The ADS scenario generation techniques mentioned above \cite{giamattei2024causality,jiang2024generation} all employ causal discovery rather than relying on the user to supply the DAG.
However, a fundamental weakness of this from a testing point of view is that inferred DAGs represent the \emph{actual system} rather than its \emph{intended behaviour}, so it will reflect any bugs in the implementation.
Causal DAGs have also been generated via static analysis of source code \cite{podgurski2020counterfault, lee2021causal}.

%% file: sections/09-conclusion.tex
\section{Conclusion}\label{sec:conclusion}
Testing nondeterministic software with uncontrollable and unobservable variables, such as ADSs, can be challenging due to the difficulty in obtaining test data.
In this paper, we investigated how two ideas from CI --- effect modification and instrumental variables --- can be used to tackle these problems.

We performed an evaluative case study by testing three requirements of the CARLA driving simulator and two associated ADSs.
Our results indicate that the above techniques can facilitate the testing of properties for which we could not otherwise obtain reliable outcomes using uncontrolled observational data.
Interaction terms in statistical estimators allow us to isolate direct causal effects in the presence of effect modification.
IV methods enable us to adjust for bias from variables that do not appear in the test data, although the accuracy will vary from system to system.
Furthermore, the main benefit of Causal Testing identified in \cite{clark2022ctf}, namely that we can obtain useful test results using observational data not collected expressly for testing, still applies in this new context.
A more extensive study to investigate the limitations and generalisability of the approach is desirable future work.

As identified in \cite{clark2022ctf,clark2023metamorphic}, the main barrier to Causal Testing is the domain knowledge necessary to draw the causal DAG.
A promising direction for future research is the creation of (semi-)automated tools to assist developers with this process.
Another line of research would be to investigate the applicability of IV methods for testing concurrent systems, where logging can hide faults \cite{helmbold1996race}.

%% file: main.bbl

\begin{thebibliography}{57}

%%% the Web address.

\ifx \showCODEN    \undefined \def \showCODEN     #1{\unskip}     \fi
\ifx \showDOI      \undefined \def \showDOI       #1{#1}\fi
\ifx \showISBNx    \undefined \def \showISBNx     #1{\unskip}     \fi
\ifx \showISBNxiii \undefined \def \showISBNxiii  #1{\unskip}     \fi
\ifx \showISSN     \undefined \def \showISSN      #1{\unskip}     \fi
\ifx \showLCCN     \undefined \def \showLCCN      #1{\unskip}     \fi
\ifx \shownote     \undefined \def \shownote      #1{#1}          \fi
\ifx \showarticletitle \undefined \def \showarticletitle #1{#1}   \fi
\ifx \showURL      \undefined \def \showURL       {\relax}        \fi
% invisible to TeX
\providecommand\bibfield[2]{#2}
\providecommand\bibinfo[2]{#2}
\providecommand\natexlab[1]{#1}
\providecommand\showeprint[2][]{arXiv:#2}

\bibitem[ped(5 05)]%
        {pedestrianissue}
 \bibinfo{year}{Accessed 2023-05-05}\natexlab{}.
\newblock \bibinfo{title}{Non determinism of Walker AI controllers}.
\newblock
\newblock
\urldef\tempurl%
\url{https://github.com/carla-simulator/carla/issues/3493}
\showURL{%
\tempurl}


\bibitem[cha(2 27)]%
        {changelog}
 \bibinfo{year}{Accessed 2024-02-27}\natexlab{}.
\newblock \bibinfo{booktitle}{\emph{CARLA changelog}}.
\newblock
\urldef\tempurl%
\url{https://github.com/carla-simulator/carla/blob/master/CHANGELOG.md}
\showURL{%
\tempurl}


\bibitem[car(9 03a)]%
        {carlaleaderboard}
 \bibinfo{year}{Accessed 2024-19-03}\natexlab{a}.
\newblock \bibinfo{title}{CARLA Autonomous Driving Leaderboard}.
\newblock
\newblock
\urldef\tempurl%
\url{https://leaderboard.carla.org}
\showURL{%
\tempurl}


\bibitem[car(9 03b)]%
        {carlagithub}
 \bibinfo{year}{Accessed 2024-19-03}\natexlab{b}.
\newblock \bibinfo{title}{CARLA simulator}.
\newblock
\newblock
\urldef\tempurl%
\url{https://github.com/carla-simulator/carla}
\showURL{%
\tempurl}


\bibitem[Arrieta et~al\mbox{.}(2021)]%
        {arrieta2021using}
\bibfield{author}{\bibinfo{person}{Aitor Arrieta}, \bibinfo{person}{Jon
  Ayerdi}, \bibinfo{person}{Miren Illarramendi}, \bibinfo{person}{Aitor
  Agirre}, \bibinfo{person}{Goiuria Sagardui}, {and} \bibinfo{person}{Maite
  Arratibel}.} \bibinfo{year}{2021}\natexlab{}.
\newblock \showarticletitle{Using machine learning to build test oracles: an
  industrial case study on elevators dispatching algorithms}. In
  \bibinfo{booktitle}{\emph{2021 IEEE/ACM International Conference on
  Automation of Software Test (AST)}}. IEEE, \bibinfo{pages}{30--39}.
\newblock


\bibitem[Baah et~al\mbox{.}(2011)]%
        {baah2011mitigating}
\bibfield{author}{\bibinfo{person}{George~K. Baah}, \bibinfo{person}{Andy
  Podgurski}, {and} \bibinfo{person}{{Mary Jean} Harrold}.}
  \bibinfo{year}{2011}\natexlab{}.
\newblock \showarticletitle{Mitigating the Confounding Effects of Program
  Dependences for Effective Fault Localization}. In
  \bibinfo{booktitle}{\emph{Proceedings of the 19th ACM SIGSOFT Symposium and
  the 13th European Conference on Foundations of Software Engineering}}
  (Szeged, Hungary) \emph{(\bibinfo{series}{ESEC/FSE '11})}.
  \bibinfo{publisher}{Association for Computing Machinery},
  \bibinfo{address}{New York, NY, USA}, \bibinfo{pages}{146–156}.
\newblock
\showISBNx{9781450304436}
\urldef\tempurl%
\url{https://doi.org/10.1145/2025113.2025136}
\showDOI{\tempurl}


\bibitem[Baah et~al\mbox{.}(2010)]%
        {baah2010causal}
\bibfield{author}{\bibinfo{person}{George~K. Baah}, \bibinfo{person}{Andy
  Podgurski}, {and} \bibinfo{person}{Mary~Jean Harrold}.}
  \bibinfo{year}{2010}\natexlab{}.
\newblock \showarticletitle{Causal Inference for Statistical Fault
  Localization}. In \bibinfo{booktitle}{\emph{Proceedings of the 19th
  International Symposium on Software Testing and Analysis}} (Trento, Italy)
  \emph{(\bibinfo{series}{ISSTA '10})}. \bibinfo{publisher}{Association for
  Computing Machinery}, \bibinfo{address}{New York, NY, USA},
  \bibinfo{pages}{73–84}.
\newblock
\showISBNx{9781605588230}
\urldef\tempurl%
\url{https://doi.org/10.1145/1831708.1831717}
\showDOI{\tempurl}


\bibitem[Bai et~al\mbox{.}(2015)]%
        {bai2015numfl}
\bibfield{author}{\bibinfo{person}{Zhuofu Bai}, \bibinfo{person}{Gang Shu},
  {and} \bibinfo{person}{Andy Podgurski}.} \bibinfo{year}{2015}\natexlab{}.
\newblock \showarticletitle{NUMFL: Localizing Faults in Numerical Software
  Using a Value-Based Causal Model}. In \bibinfo{booktitle}{\emph{2015 IEEE 8th
  International Conference on Software Testing, Verification and Validation
  (ICST)}}. \bibinfo{publisher}{IEEE}, \bibinfo{pages}{1--10}.
\newblock
\urldef\tempurl%
\url{https://doi.org/10.1109/ICST.2015.7102597}
\showDOI{\tempurl}


\bibitem[Bareinboim and Pearl(2016)]%
        {bareinboim2016causal}
\bibfield{author}{\bibinfo{person}{Elias Bareinboim} {and}
  \bibinfo{person}{Judea Pearl}.} \bibinfo{year}{2016}\natexlab{}.
\newblock \showarticletitle{Causal inference and the data-fusion problem}.
\newblock \bibinfo{journal}{\emph{Proceedings of the National Academy of
  Sciences of the United States of America}} \bibinfo{volume}{113},
  \bibinfo{number}{27} (\bibinfo{year}{2016}), \bibinfo{pages}{7345--7352}.
\newblock
\showISSN{00278424, 10916490}


\bibitem[Barr et~al\mbox{.}(2015)]%
        {barr2015oracle}
\bibfield{author}{\bibinfo{person}{Earl~T. Barr}, \bibinfo{person}{Mark
  Harman}, \bibinfo{person}{Phil McMinn}, \bibinfo{person}{Muzammil Shahbaz},
  {and} \bibinfo{person}{Shin Yoo}.} \bibinfo{year}{2015}\natexlab{}.
\newblock \showarticletitle{The Oracle Problem in Software Testing: A Survey}.
\newblock \bibinfo{journal}{\emph{IEEE Transactions on Software Engineering}}
  \bibinfo{volume}{41}, \bibinfo{number}{5} (\bibinfo{year}{2015}),
  \bibinfo{pages}{507--525}.
\newblock
\urldef\tempurl%
\url{https://doi.org/10.1109/TSE.2014.2372785}
\showDOI{\tempurl}


\bibitem[Chen and Jiang(2021)]%
        {chen2021logs}
\bibfield{author}{\bibinfo{person}{Boyuan Chen} {and} \bibinfo{person}{Zhen
  Ming~(Jack) Jiang}.} \bibinfo{year}{2021}\natexlab{}.
\newblock \showarticletitle{A Survey of Software Log Instrumentation}.
\newblock \bibinfo{journal}{\emph{Comput. Surveys}} \bibinfo{volume}{54},
  \bibinfo{number}{4} (\bibinfo{date}{may} \bibinfo{year}{2021}),
  \bibinfo{pages}{1–34}.
\newblock
\showISSN{1557-7341}
\urldef\tempurl%
\url{https://doi.org/10.1145/3448976}
\showDOI{\tempurl}


\bibitem[Chen et~al\mbox{.}(1998)]%
        {chen1998metamorphic}
\bibfield{author}{\bibinfo{person}{Tsong~Y. Chen}, \bibinfo{person}{Shing~C.
  Cheung}, {and} \bibinfo{person}{Shiu~Ming Yiu}.}
  \bibinfo{year}{1998}\natexlab{}.
\newblock \bibinfo{booktitle}{\emph{Metamorphic testing: A new approach for
  generating next test cases}}.
\newblock \bibinfo{type}{{T}echnical {R}eport} HKUST-CS98-01.
  \bibinfo{institution}{The Hong Kong University of Science and Technology}.
\newblock


\bibitem[Chen et~al\mbox{.}(2020)]%
        {chen2020learning}
\bibfield{author}{\bibinfo{person}{Yuqi Chen}, \bibinfo{person}{Christopher~M.
  Poskitt}, \bibinfo{person}{Jun Sun}, \bibinfo{person}{Sridhar Adepu}, {and}
  \bibinfo{person}{Fan Zhang}.} \bibinfo{year}{2020}\natexlab{}.
\newblock \showarticletitle{Learning-guided network fuzzing for testing
  cyber-physical system defences}. In \bibinfo{booktitle}{\emph{Proceedings of
  the 34th IEEE/ACM International Conference on Automated Software
  Engineering}} \emph{(\bibinfo{series}{ASE '19})}. \bibinfo{publisher}{IEEE
  Press}, \bibinfo{pages}{962–973}.
\newblock
\showISBNx{9781728125084}
\urldef\tempurl%
\url{https://doi.org/10.1109/ASE.2019.00093}
\showDOI{\tempurl}


\bibitem[Cheng and Krishnakumar(1993)]%
        {chen1995efsm}
\bibfield{author}{\bibinfo{person}{Kwang~Ting Cheng} {and}
  \bibinfo{person}{A.~S. Krishnakumar}.} \bibinfo{year}{1993}\natexlab{}.
\newblock \showarticletitle{Automatic functional test generation using the
  extended finite state machine model}. In
  \bibinfo{booktitle}{\emph{Proceedings of the 30th International Design
  Automation Conference}} (Dallas, Texas, USA) \emph{(\bibinfo{series}{DAC
  '93})}. \bibinfo{publisher}{Association for Computing Machinery},
  \bibinfo{address}{New York, NY, USA}, \bibinfo{pages}{86–91}.
\newblock
\showISBNx{0897915771}
\urldef\tempurl%
\url{https://doi.org/10.1145/157485.164585}
\showDOI{\tempurl}


\bibitem[Clark et~al\mbox{.}(2023a)]%
        {clark2022ctf}
\bibfield{author}{\bibinfo{person}{Andrew~G. Clark}, \bibinfo{person}{Michael
  Foster}, \bibinfo{person}{Benedikt Prifling}, \bibinfo{person}{Neil
  Walkinshaw}, \bibinfo{person}{Robert~M. Hierons}, \bibinfo{person}{Volker
  Schmidt}, {and} \bibinfo{person}{Robert~D. Turner}.}
  \bibinfo{year}{2023}\natexlab{a}.
\newblock \showarticletitle{Testing Causality in Scientific Modelling
  Software}.
\newblock \bibinfo{journal}{\emph{ACM Trans. Softw. Eng. Methodol.}}
  \bibinfo{volume}{33}, \bibinfo{number}{1}, Article \bibinfo{articleno}{10}
  (\bibinfo{date}{nov} \bibinfo{year}{2023}), \bibinfo{numpages}{42}~pages.
\newblock
\showISSN{1049-331X}
\urldef\tempurl%
\url{https://doi.org/10.1145/3607184}
\showDOI{\tempurl}


\bibitem[Clark et~al\mbox{.}(2023b)]%
        {clark2023metamorphic}
\bibfield{author}{\bibinfo{person}{Andrew~G. Clark}, \bibinfo{person}{Michael
  Foster}, \bibinfo{person}{Neil Walkinshaw}, {and} \bibinfo{person}{Robert~M.
  Hierons}.} \bibinfo{year}{2023}\natexlab{b}.
\newblock \showarticletitle{Metamorphic Testing with Causal Graphs}. In
  \bibinfo{booktitle}{\emph{2023 IEEE Conference on Software Testing,
  Verification and Validation (ICST)}}. \bibinfo{pages}{153--164}.
\newblock
\urldef\tempurl%
\url{https://doi.org/10.1109/ICST57152.2023.00023}
\showDOI{\tempurl}


\bibitem[Dosovitskiy et~al\mbox{.}(2017)]%
        {dosovitskiy2017carla}
\bibfield{author}{\bibinfo{person}{Alexey Dosovitskiy}, \bibinfo{person}{German
  Ros}, \bibinfo{person}{Felipe Codevilla}, \bibinfo{person}{Antonio Lopez},
  {and} \bibinfo{person}{Vladlen Koltun}.} \bibinfo{year}{2017}\natexlab{}.
\newblock \showarticletitle{{CARLA}: {An} Open Urban Driving Simulator}. In
  \bibinfo{booktitle}{\emph{Proceedings of the 1st Annual Conference on Robot
  Learning}} \emph{(\bibinfo{series}{Proceedings of Machine Learning Research},
  Vol.~\bibinfo{volume}{78})}. \bibinfo{publisher}{PMLR},
  \bibinfo{pages}{1--16}.
\newblock


\bibitem[Foster et~al\mbox{.}(2024)]%
        {foster2024adequacy}
\bibfield{author}{\bibinfo{person}{Michael Foster},
  \bibinfo{person}{Christopher Wild}, \bibinfo{person}{Robert~M. Hierons},
  {and} \bibinfo{person}{Neil Walkinshaw}.} \bibinfo{year}{2024}\natexlab{}.
\newblock \showarticletitle{Causal Test Adequacy}. In
  \bibinfo{booktitle}{\emph{2024 IEEE Conference on Software Testing,
  Verification and Validation (ICST)}}. \bibinfo{publisher}{IEEE},
  \bibinfo{pages}{161–172}.
\newblock
\urldef\tempurl%
\url{https://doi.org/10.1109/icst60714.2024.00023}
\showDOI{\tempurl}


\bibitem[Freedman(1991)]%
        {freedman1991testability}
\bibfield{author}{\bibinfo{person}{R.S. Freedman}.}
  \bibinfo{year}{1991}\natexlab{}.
\newblock \showarticletitle{Testability of software components}.
\newblock \bibinfo{journal}{\emph{IEEE Transactions on Software Engineering}}
  \bibinfo{volume}{17}, \bibinfo{number}{6} (\bibinfo{year}{1991}),
  \bibinfo{pages}{553--564}.
\newblock
\urldef\tempurl%
\url{https://doi.org/10.1109/32.87281}
\showDOI{\tempurl}


\bibitem[Furia et~al\mbox{.}(2023)]%
        {furia2023towards}
\bibfield{author}{\bibinfo{person}{Carlo~A Furia}, \bibinfo{person}{Richard
  Torkar}, {and} \bibinfo{person}{Robert Feldt}.}
  \bibinfo{year}{2023}\natexlab{}.
\newblock \showarticletitle{Towards causal analysis of empirical software
  engineering data: The impact of programming languages on coding
  competitions}.
\newblock \bibinfo{journal}{\emph{ACM Transactions on Software Engineering and
  Methodology}} \bibinfo{volume}{33}, \bibinfo{number}{1}
  (\bibinfo{year}{2023}), \bibinfo{pages}{1--35}.
\newblock


\bibitem[Giamattei et~al\mbox{.}(2024)]%
        {giamattei2024causality}
\bibfield{author}{\bibinfo{person}{Luca Giamattei}, \bibinfo{person}{Antonio
  Guerriero}, \bibinfo{person}{Roberto Pietrantuono}, {and}
  \bibinfo{person}{Stefano Russo}.} \bibinfo{year}{2024}\natexlab{}.
\newblock \showarticletitle{Causality-driven Testing of Autonomous Driving
  Systems}.
\newblock \bibinfo{journal}{\emph{ACM Trans. Softw. Eng. Methodol.}}
  \bibinfo{volume}{33}, \bibinfo{number}{3}, Article \bibinfo{articleno}{74}
  (\bibinfo{year}{2024}), \bibinfo{numpages}{35}~pages.
\newblock
\showISSN{1049-331X}
\urldef\tempurl%
\url{https://doi.org/10.1145/3635709}
\showDOI{\tempurl}


\bibitem[Giamattei et~al\mbox{.}(2025)]%
        {giamattei2025survey}
\bibfield{author}{\bibinfo{person}{Luca Giamattei}, \bibinfo{person}{Antonio
  Guerriero}, \bibinfo{person}{Roberto Pietrantuono}, {and}
  \bibinfo{person}{Stefano Russo}.} \bibinfo{year}{2025}\natexlab{}.
\newblock \showarticletitle{Causal reasoning in Software Quality Assurance: A
  systematic review}.
\newblock \bibinfo{journal}{\emph{Information and Software Technology}}
  \bibinfo{volume}{178} (\bibinfo{date}{Feb.} \bibinfo{year}{2025}),
  \bibinfo{pages}{107599}.
\newblock
\showISSN{0950-5849}
\urldef\tempurl%
\url{https://doi.org/10.1016/j.infsof.2024.107599}
\showDOI{\tempurl}


\bibitem[Glymour et~al\mbox{.}(2019)]%
        {glymour2019review}
\bibfield{author}{\bibinfo{person}{Clark Glymour}, \bibinfo{person}{Kun Zhang},
  {and} \bibinfo{person}{Peter Spirtes}.} \bibinfo{year}{2019}\natexlab{}.
\newblock \showarticletitle{Review of causal discovery methods based on
  graphical models}.
\newblock \bibinfo{journal}{\emph{Frontiers in genetics}}  \bibinfo{volume}{10}
  (\bibinfo{year}{2019}), \bibinfo{pages}{524}.
\newblock


\bibitem[Gore and Reynolds(2012)]%
        {gore2012reducing}
\bibfield{author}{\bibinfo{person}{Ross Gore} {and} \bibinfo{person}{Paul~F.
  Reynolds}.} \bibinfo{year}{2012}\natexlab{}.
\newblock \showarticletitle{Reducing confounding bias in predicate-level
  statistical debugging metrics}. In \bibinfo{booktitle}{\emph{2012 34th
  International Conference on Software Engineering (ICSE)}}.
  \bibinfo{publisher}{IEEE}, \bibinfo{pages}{463--473}.
\newblock
\urldef\tempurl%
\url{https://doi.org/10.1109/ICSE.2012.6227169}
\showDOI{\tempurl}


\bibitem[Guderlei and Mayer(2007)]%
        {guderlei2007statistical}
\bibfield{author}{\bibinfo{person}{Ralph Guderlei} {and}
  \bibinfo{person}{Johannes Mayer}.} \bibinfo{year}{2007}\natexlab{}.
\newblock \showarticletitle{Statistical Metamorphic Testing Testing Programs
  with Random Output by Means of Statistical Hypothesis Tests and Metamorphic
  Testing}. In \bibinfo{booktitle}{\emph{Seventh International Conference on
  Quality Software (QSIC 2007)}}. \bibinfo{pages}{404--409}.
\newblock
\urldef\tempurl%
\url{https://doi.org/10.1109/QSIC.2007.4385527}
\showDOI{\tempurl}


\bibitem[Han and Zhou(2020)]%
        {han2020fuzz}
\bibfield{author}{\bibinfo{person}{Jia~Cheng Han} {and}
  \bibinfo{person}{Zhi~Quan Zhou}.} \bibinfo{year}{2020}\natexlab{}.
\newblock \showarticletitle{Metamorphic Fuzz Testing of Autonomous Vehicles}.
  In \bibinfo{booktitle}{\emph{Proceedings of the IEEE/ACM 42nd International
  Conference on Software Engineering Workshops}} (Seoul, Republic of Korea)
  \emph{(\bibinfo{series}{ICSEW'20})}. \bibinfo{publisher}{Association for
  Computing Machinery}, \bibinfo{address}{New York, NY, USA},
  \bibinfo{pages}{380–385}.
\newblock
\showISBNx{9781450379632}
\urldef\tempurl%
\url{https://doi.org/10.1145/3387940.3392252}
\showDOI{\tempurl}


\bibitem[Haq et~al\mbox{.}(2022)]%
        {haq2022efficient}
\bibfield{author}{\bibinfo{person}{Fitash~Ul Haq}, \bibinfo{person}{Donghwan
  Shin}, {and} \bibinfo{person}{Lionel Briand}.}
  \bibinfo{year}{2022}\natexlab{}.
\newblock \showarticletitle{Efficient Online Testing for DNN-Enabled Systems
  Using Surrogate-Assisted and Many-Objective Optimization}. In
  \bibinfo{booktitle}{\emph{Proceedings of the 44th International Conference on
  Software Engineering}} \emph{(\bibinfo{series}{ICSE '22})}.
  \bibinfo{publisher}{Association for Computing Machinery},
  \bibinfo{address}{New York, NY, USA}, \bibinfo{pages}{811–822}.
\newblock
\showISBNx{9781450392211}
\urldef\tempurl%
\url{https://doi.org/10.1145/3510003.3510188}
\showDOI{\tempurl}


\bibitem[Helmbold and McDowell(1996)]%
        {helmbold1996race}
\bibfield{author}{\bibinfo{person}{D.P. Helmbold} {and} \bibinfo{person}{C.E.
  McDowell}.} \bibinfo{year}{1996}\natexlab{}.
\newblock \showarticletitle{A Taxonomy of Race Conditions}.
\newblock \bibinfo{journal}{\emph{J. Parallel and Distrib. Comput.}}
  \bibinfo{volume}{33}, \bibinfo{number}{2} (\bibinfo{year}{1996}),
  \bibinfo{pages}{159--164}.
\newblock
\showISSN{0743-7315}
\urldef\tempurl%
\url{https://doi.org/10.1006/jpdc.1996.0034}
\showDOI{\tempurl}


\bibitem[Hern{\'a}n and Robins(2020)]%
        {hernan2020causal}
\bibfield{author}{\bibinfo{person}{Miguel~A Hern{\'a}n} {and}
  \bibinfo{person}{James~M Robins}.} \bibinfo{year}{2020}\natexlab{}.
\newblock \bibinfo{booktitle}{\emph{Causal {I}nference: {What} if}}.
\newblock \bibinfo{publisher}{Chapman \& Hall/CRC}, \bibinfo{address}{Boca
  Raton}.
\newblock


\bibitem[Jaeger et~al\mbox{.}(2023)]%
        {jaeger2023garage}
\bibfield{author}{\bibinfo{person}{Bernhard Jaeger}, \bibinfo{person}{Kashyap
  Chitta}, {and} \bibinfo{person}{Andreas Geiger}.}
  \bibinfo{year}{2023}\natexlab{}.
\newblock \showarticletitle{Hidden Biases of End-to-End Driving Models}. In
  \bibinfo{booktitle}{\emph{2023 IEEE/CVF International Conference on Computer
  Vision (ICCV)}}. \bibinfo{publisher}{IEEE}.
\newblock
\urldef\tempurl%
\url{https://doi.org/10.1109/iccv51070.2023.00757}
\showDOI{\tempurl}


\bibitem[Jiang et~al\mbox{.}(2024)]%
        {jiang2024generation}
\bibfield{author}{\bibinfo{person}{Zhengmin Jiang}, \bibinfo{person}{Jia Liu},
  \bibinfo{person}{Peng Sun}, \bibinfo{person}{Ming Sang},
  \bibinfo{person}{Huiyun Li}, {and} \bibinfo{person}{Yi Pan}.}
  \bibinfo{year}{2024}\natexlab{}.
\newblock \showarticletitle{Generation of Risky Scenarios for Testing Automated
  Driving Visual Perception Based on Causal Analysis}.
\newblock \bibinfo{journal}{\emph{IEEE Transactions on Intelligent
  Transportation Systems}} \bibinfo{volume}{25}, \bibinfo{number}{11}
  (\bibinfo{year}{2024}), \bibinfo{pages}{15991--16004}.
\newblock
\urldef\tempurl%
\url{https://doi.org/10.1109/TITS.2024.3421343}
\showDOI{\tempurl}


\bibitem[Johnson et~al\mbox{.}(2020)]%
        {johnson2020causal}
\bibfield{author}{\bibinfo{person}{Brittany Johnson}, \bibinfo{person}{Yuriy
  Brun}, {and} \bibinfo{person}{Alexandra Meliou}.}
  \bibinfo{year}{2020}\natexlab{}.
\newblock \showarticletitle{Causal testing: understanding defects' root
  causes}. In \bibinfo{booktitle}{\emph{Proceedings of the ACM/IEEE 42nd
  International Conference on Software Engineering}}. \bibinfo{pages}{87--99}.
\newblock


\bibitem[Kaur et~al\mbox{.}(2021)]%
        {kaur2021simulators}
\bibfield{author}{\bibinfo{person}{Prabhjot Kaur}, \bibinfo{person}{Samira
  Taghavi}, \bibinfo{person}{Zhaofeng Tian}, {and} \bibinfo{person}{Weisong
  Shi}.} \bibinfo{year}{2021}\natexlab{}.
\newblock \showarticletitle{A Survey on Simulators for Testing Self-Driving
  Cars}. In \bibinfo{booktitle}{\emph{2021 Fourth International Conference on
  Connected and Autonomous Driving (MetroCAD)}}. \bibinfo{pages}{62--70}.
\newblock
\urldef\tempurl%
\url{https://doi.org/10.1109/MetroCAD51599.2021.00018}
\showDOI{\tempurl}


\bibitem[Lee et~al\mbox{.}(2021)]%
        {lee2021causal}
\bibfield{author}{\bibinfo{person}{Seongmin Lee}, \bibinfo{person}{Dave
  Binkley}, \bibinfo{person}{Robert Feldt}, \bibinfo{person}{Nicolas Gold},
  {and} \bibinfo{person}{Shin Yoo}.} \bibinfo{year}{2021}\natexlab{}.
\newblock \showarticletitle{Causal program dependence analysis}.
\newblock \bibinfo{journal}{\emph{arXiv preprint arXiv:2104.09107}}
  (\bibinfo{year}{2021}).
\newblock


\bibitem[Malinsky and Danks(2018)]%
        {malinsky2018causal}
\bibfield{author}{\bibinfo{person}{Daniel Malinsky} {and}
  \bibinfo{person}{David Danks}.} \bibinfo{year}{2018}\natexlab{}.
\newblock \showarticletitle{Causal discovery algorithms: A practical guide}.
\newblock \bibinfo{journal}{\emph{Philosophy Compass}} \bibinfo{volume}{13},
  \bibinfo{number}{1} (\bibinfo{year}{2018}), \bibinfo{pages}{e12470}.
\newblock


\bibitem[McConnell and Lindner(2019)]%
        {mcConnell2019estimating}
\bibfield{author}{\bibinfo{person}{K.~John McConnell} {and}
  \bibinfo{person}{Stephan Lindner}.} \bibinfo{year}{2019}\natexlab{}.
\newblock \showarticletitle{Estimating treatment effects with machine
  learning}.
\newblock \bibinfo{journal}{\emph{Health Services Research}}
  \bibinfo{volume}{54}, \bibinfo{number}{6} (\bibinfo{date}{oct}
  \bibinfo{year}{2019}), \bibinfo{pages}{1273–1282}.
\newblock
\showISSN{1475-6773}
\urldef\tempurl%
\url{https://doi.org/10.1111/1475-6773.13212}
\showDOI{\tempurl}


\bibitem[Nie and Leung(2011)]%
        {nie2011combinatorial}
\bibfield{author}{\bibinfo{person}{Changhai Nie} {and} \bibinfo{person}{Hareton
  Leung}.} \bibinfo{year}{2011}\natexlab{}.
\newblock \showarticletitle{A survey of combinatorial testing}.
\newblock \bibinfo{journal}{\emph{ACM Comput. Surv.}} \bibinfo{volume}{43},
  \bibinfo{number}{2}, Article \bibinfo{articleno}{11} (\bibinfo{date}{feb}
  \bibinfo{year}{2011}), \bibinfo{numpages}{29}~pages.
\newblock
\showISSN{0360-0300}
\urldef\tempurl%
\url{https://doi.org/10.1145/1883612.1883618}
\showDOI{\tempurl}


\bibitem[Nitsche et~al\mbox{.}(2018)]%
        {nitsche2018junctions}
\bibfield{author}{\bibinfo{person}{P. Nitsche}, \bibinfo{person}{R.H. Welsh},
  \bibinfo{person}{A. Genser}, {and} \bibinfo{person}{P.D. Thomas}.}
  \bibinfo{year}{2018}\natexlab{}.
\newblock \showarticletitle{A novel, modular validation framework for collision
  avoidance of automated vehicles at road junctions}. In
  \bibinfo{booktitle}{\emph{2018 21st International Conference on Intelligent
  Transportation Systems (ITSC)}}. \bibinfo{pages}{90--97}.
\newblock
\urldef\tempurl%
\url{https://doi.org/10.1109/ITSC.2018.8569631}
\showDOI{\tempurl}


\bibitem[Norden et~al\mbox{.}(2019)]%
        {norden2019estimation}
\bibfield{author}{\bibinfo{person}{Justin Norden}, \bibinfo{person}{Matthew
  O'Kelly}, {and} \bibinfo{person}{Aman Sinha}.}
  \bibinfo{year}{2019}\natexlab{}.
\newblock \bibinfo{title}{Efficient Black-box Assessment of Autonomous Vehicle
  Safety}.
\newblock
\newblock
\urldef\tempurl%
\url{https://doi.org/10.48550/ARXIV.1912.03618}
\showDOI{\tempurl}


\bibitem[O'Brien and Yi(2016)]%
        {o2016interpret}
\bibfield{author}{\bibinfo{person}{Sheila~F O'Brien} {and}
  \bibinfo{person}{Qi~Long Yi}.} \bibinfo{year}{2016}\natexlab{}.
\newblock \showarticletitle{How do I interpret a confidence interval?}
\newblock \bibinfo{journal}{\emph{Transfusion}} \bibinfo{volume}{56},
  \bibinfo{number}{7} (\bibinfo{year}{2016}), \bibinfo{pages}{1680--1683}.
\newblock


\bibitem[Pearl(2009)]%
        {pearl2009causality}
\bibfield{author}{\bibinfo{person}{Judea Pearl}.}
  \bibinfo{year}{2009}\natexlab{}.
\newblock \bibinfo{booktitle}{\emph{Causality}}.
\newblock \bibinfo{publisher}{Cambridge university press},
  \bibinfo{address}{Cambridge}.
\newblock
\showISBNx{9780521895606}


\bibitem[Podgurski and K{\"u}{\c{c}}{\"u}k(2020)]%
        {podgurski2020counterfault}
\bibfield{author}{\bibinfo{person}{Andy Podgurski} {and}
  \bibinfo{person}{Yi{\u{g}}it K{\"u}{\c{c}}{\"u}k}.}
  \bibinfo{year}{2020}\natexlab{}.
\newblock \showarticletitle{CounterFault: Value-Based Fault Localization by
  Modeling and Predicting Counterfactual Outcomes}. In
  \bibinfo{booktitle}{\emph{2020 IEEE International Conference on Software
  Maintenance and Evolution (ICSME)}}. IEEE, \bibinfo{pages}{382--393}.
\newblock


\bibitem[Poskitt et~al\mbox{.}(2023)]%
        {poskitt2023}
\bibfield{author}{\bibinfo{person}{Christopher~M. Poskitt},
  \bibinfo{person}{Yuqi Chen}, \bibinfo{person}{Jun Sun}, {and}
  \bibinfo{person}{Yu Jiang}.} \bibinfo{year}{2023}\natexlab{}.
\newblock \showarticletitle{Finding Causally Different Tests for an Industrial
  Control System}. In \bibinfo{booktitle}{\emph{2023 IEEE/ACM 45th
  International Conference on Software Engineering (ICSE)}}.
  \bibinfo{pages}{2578--2590}.
\newblock
\urldef\tempurl%
\url{https://doi.org/10.1109/ICSE48619.2023.00215}
\showDOI{\tempurl}


\bibitem[{Ralph et al.}(2021)]%
        {ralph2021empirical}
\bibfield{author}{\bibinfo{person}{Paul {Ralph et al.}}}
  \bibinfo{year}{2021}\natexlab{}.
\newblock \bibinfo{title}{Empirical Standards for Software Engineering
  Research}.
\newblock
\newblock
\showeprint[arxiv]{2010.03525}~[cs.SE]


\bibitem[Runeson et~al\mbox{.}(2012)]%
        {runeson2012case}
\bibfield{author}{\bibinfo{person}{Per Runeson}, \bibinfo{person}{Martin Host},
  \bibinfo{person}{Austen Rainer}, {and} \bibinfo{person}{Bjorn Regnell}.}
  \bibinfo{year}{2012}\natexlab{}.
\newblock \bibinfo{booktitle}{\emph{Case study research in software
  engineering: Guidelines and examples}}.
\newblock \bibinfo{publisher}{John Wiley \& Sons}.
\newblock


\bibitem[Shu et~al\mbox{.}(2013)]%
        {shu2013mfl}
\bibfield{author}{\bibinfo{person}{Gang Shu}, \bibinfo{person}{Boya Sun},
  \bibinfo{person}{Andy Podgurski}, {and} \bibinfo{person}{Feng Cao}.}
  \bibinfo{year}{2013}\natexlab{}.
\newblock \showarticletitle{Mfl: Method-level fault localization with causal
  inference}. In \bibinfo{booktitle}{\emph{2013 IEEE Sixth International
  Conference on Software Testing, Verification and Validation}}. IEEE,
  \bibinfo{publisher}{IEEE}, \bibinfo{pages}{124--133}.
\newblock


\bibitem[Siebert(2023)]%
        {siebert2023applications}
\bibfield{author}{\bibinfo{person}{Julien Siebert}.}
  \bibinfo{year}{2023}\natexlab{}.
\newblock \showarticletitle{Applications of statistical causal inference in
  software engineering}.
\newblock \bibinfo{journal}{\emph{Information and Software Technology}}
  \bibinfo{volume}{159} (\bibinfo{date}{jul} \bibinfo{year}{2023}),
  \bibinfo{pages}{107198}.
\newblock
\showISSN{0950-5849}
\urldef\tempurl%
\url{https://doi.org/10.1016/j.infsof.2023.107198}
\showDOI{\tempurl}


\bibitem[Siroker and Koomen(2015)]%
        {siroker2015abtesting}
\bibfield{author}{\bibinfo{person}{Dan Siroker} {and} \bibinfo{person}{Pete
  Koomen}.} \bibinfo{year}{2015}\natexlab{}.
\newblock \bibinfo{booktitle}{\emph{A/B testing: The most powerful way to turn
  clicks into customers}}.
\newblock \bibinfo{publisher}{John Wiley \& Sons}.
\newblock
\showISBNx{9781118536094}


\bibitem[Sun et~al\mbox{.}(2022)]%
        {sun2022scenario}
\bibfield{author}{\bibinfo{person}{Jian Sun}, \bibinfo{person}{He Zhang},
  \bibinfo{person}{Huajun Zhou}, \bibinfo{person}{Rongjie Yu}, {and}
  \bibinfo{person}{Ye Tian}.} \bibinfo{year}{2022}\natexlab{}.
\newblock \showarticletitle{Scenario-Based Test Automation for Highly Automated
  Vehicles: A Review and Paving the Way for Systematic Safety Assurance}.
\newblock \bibinfo{journal}{\emph{IEEE Transactions on Intelligent
  Transportation Systems}} \bibinfo{volume}{23}, \bibinfo{number}{9}
  (\bibinfo{year}{2022}), \bibinfo{pages}{14088--14103}.
\newblock
\urldef\tempurl%
\url{https://doi.org/10.1109/TITS.2021.3136353}
\showDOI{\tempurl}


\bibitem[Tang et~al\mbox{.}(2023)]%
        {tang2023survey}
\bibfield{author}{\bibinfo{person}{Shuncheng Tang}, \bibinfo{person}{Zhenya
  Zhang}, \bibinfo{person}{Yi Zhang}, \bibinfo{person}{Jixiang Zhou},
  \bibinfo{person}{Yan Guo}, \bibinfo{person}{Shuang Liu},
  \bibinfo{person}{Shengjian Guo}, \bibinfo{person}{Yan-Fu Li},
  \bibinfo{person}{Lei Ma}, \bibinfo{person}{Yinxing Xue}, {and}
  \bibinfo{person}{Yang Liu}.} \bibinfo{year}{2023}\natexlab{}.
\newblock \showarticletitle{A Survey on Automated Driving System Testing:
  Landscapes and Trends}.
\newblock \bibinfo{journal}{\emph{ACM Trans. Softw. Eng. Methodol.}}
  \bibinfo{volume}{32}, \bibinfo{number}{5}, Article \bibinfo{articleno}{124}
  (\bibinfo{date}{jul} \bibinfo{year}{2023}), \bibinfo{numpages}{62}~pages.
\newblock
\showISSN{1049-331X}
\urldef\tempurl%
\url{https://doi.org/10.1145/3579642}
\showDOI{\tempurl}


\bibitem[V.~Basili and Rombach(1994)]%
        {basili1994encyclopedia}
\bibfield{author}{\bibinfo{person}{C.~Caldiera V.~Basili} {and}
  \bibinfo{person}{D.~H. Rombach}.} \bibinfo{year}{1994}\natexlab{}.
\newblock \bibinfo{booktitle}{\emph{Goal question metric paradigm}}.
  Vol.~\bibinfo{volume}{2}.
\newblock \bibinfo{publisher}{Wiley}. 528--532 pages.
\newblock


\bibitem[Weinberg(2007)]%
        {weinberg2007modification}
\bibfield{author}{\bibinfo{person}{Clarice~R. Weinberg}.}
  \bibinfo{year}{2007}\natexlab{}.
\newblock \showarticletitle{Can {DAGs} Clarify Effect Modification?}
\newblock \bibinfo{journal}{\emph{Epidemiology}} \bibinfo{volume}{18},
  \bibinfo{number}{5} (\bibinfo{date}{sep} \bibinfo{year}{2007}),
  \bibinfo{pages}{569–572}.
\newblock
\showISSN{1044-3983}
\urldef\tempurl%
\url{https://doi.org/10.1097/ede.0b013e318126c11d}
\showDOI{\tempurl}


\bibitem[Wright(1920)]%
        {wright1920relative}
\bibfield{author}{\bibinfo{person}{S Wright}.} \bibinfo{year}{1920}\natexlab{}.
\newblock \showarticletitle{The Relative Importance of Heredity and Environment
  in Determining the Piebald Pattern of {Guinea-Pigs}}.
\newblock \bibinfo{journal}{\emph{Proc Natl Acad Sci U S A}}
  \bibinfo{volume}{6}, \bibinfo{number}{6} (\bibinfo{date}{jun}
  \bibinfo{year}{1920}), \bibinfo{pages}{320--332}.
\newblock


\bibitem[Wu et~al\mbox{.}(2022)]%
        {wu2022TCP}
\bibfield{author}{\bibinfo{person}{Penghao Wu}, \bibinfo{person}{Xiaosong Jia},
  \bibinfo{person}{Li Chen}, \bibinfo{person}{Junchi Yan},
  \bibinfo{person}{Hongyang Li}, {and} \bibinfo{person}{Yu Qiao}.}
  \bibinfo{year}{2022}\natexlab{}.
\newblock \bibinfo{title}{Trajectory-guided Control Prediction for End-to-end
  Autonomous Driving: A Simple yet Strong Baseline}.
\newblock
\newblock
\showeprint[arxiv]{2206.08129}~[cs.CV]


\bibitem[Zhang et~al\mbox{.}(2023)]%
        {zhang2023finding}
\bibfield{author}{\bibinfo{person}{Xinhai Zhang}, \bibinfo{person}{Jianbo Tao},
  \bibinfo{person}{Kaige Tan}, \bibinfo{person}{Martin Törngren},
  \bibinfo{person}{José Manuel~Gaspar Sánchez},
  \bibinfo{person}{Muhammad~Rusyadi Ramli}, \bibinfo{person}{Xin Tao},
  \bibinfo{person}{Magnus Gyllenhammar}, \bibinfo{person}{Franz Wotawa},
  \bibinfo{person}{Naveen Mohan}, \bibinfo{person}{Mihai Nica}, {and}
  \bibinfo{person}{Hermann Felbinger}.} \bibinfo{year}{2023}\natexlab{}.
\newblock \showarticletitle{Finding Critical Scenarios for Automated Driving
  Systems: A Systematic Mapping Study}.
\newblock \bibinfo{journal}{\emph{IEEE Transactions on Software Engineering}}
  \bibinfo{volume}{49}, \bibinfo{number}{3} (\bibinfo{year}{2023}),
  \bibinfo{pages}{991--1026}.
\newblock
\urldef\tempurl%
\url{https://doi.org/10.1109/TSE.2022.3170122}
\showDOI{\tempurl}


\bibitem[Zhong et~al\mbox{.}(2021)]%
        {zhong2021survey}
\bibfield{author}{\bibinfo{person}{Ziyuan Zhong}, \bibinfo{person}{Yun Tang},
  \bibinfo{person}{Yuan Zhou}, \bibinfo{person}{Vania de~Oliveira Neves},
  \bibinfo{person}{Yang Liu}, {and} \bibinfo{person}{Baishakhi Ray}.}
  \bibinfo{year}{2021}\natexlab{}.
\newblock \bibinfo{title}{A Survey on Scenario-Based Testing for Automated
  Driving Systems in High-Fidelity Simulation}.
\newblock
\newblock
\urldef\tempurl%
\url{https://doi.org/10.48550/ARXIV.2112.00964}
\showDOI{\tempurl}


\bibitem[Zohdinasab et~al\mbox{.}(2023)]%
        {deephyperion-cs}
\bibfield{author}{\bibinfo{person}{Tahereh Zohdinasab},
  \bibinfo{person}{Vincenzo Riccio}, \bibinfo{person}{Alessio Gambi}, {and}
  \bibinfo{person}{Paolo Tonella}.} \bibinfo{year}{2023}\natexlab{}.
\newblock \showarticletitle{Efficient and Effective Feature Space Exploration
  for Testing Deep Learning Systems}.
\newblock \bibinfo{journal}{\emph{ACM Trans. Softw. Eng. Methodol.}}
  \bibinfo{volume}{32}, \bibinfo{number}{2}, Article \bibinfo{articleno}{49}
  (\bibinfo{date}{mar} \bibinfo{year}{2023}), \bibinfo{numpages}{38}~pages.
\newblock
\showISSN{1049-331X}
\urldef\tempurl%
\url{https://doi.org/10.1145/3544792}
\showDOI{\tempurl}


\end{thebibliography}
